\documentclass[%
 reprint,
 amsmath,amssymb,
 aps,pra
]{revtex4-2}

\usepackage{graphicx}
\usepackage{amsmath,mathtools}
\usepackage{tikz}
\usepackage{caption}
\usepackage[normalem]{ulem}

\newcommand{\addtxt}[1]{{\textcolor{black}{#1}}}

\begin{document}


\title{Benchmarking and contrasting exchange-correlation functional differences in response to static correlation in unrestricted Kohn-Sham and \addtxt{a hybrid} 1-electron reduced density matrix functional theory}

\author{Daniel Gibney}
\author{Jan-Niklas Boyn}%
\email{jboyn@umn.edu}%
\affiliation{Department of Chemistry, University of Minnesota, Minneapolis, Minnesota 55455, United States}%
\date{\today}

\begin{abstract}
A hybrid Kohn-Sham Density Functional Theory (KS-DFT) \addtxt{and} 1-electron Reduced Density Matrix Functional Theory (1-RDMFT) has recently been developed to describe strongly correlated systems at mean-field computational cost. This approach relies on combining a Reduced Density Matrix Functional to capture strong correlation effects with existing exchange correlation (XC) functionals to capture the remaining dynamical correlation effects. In this work, we systematically benchmark the performance of nearly 200 different XC functionals available within LibXC in this \addtxt{DFA} 1-RDMFT framework, contrasting it with their performance in unrestricted KS-DFT. We identify optimal XC functionals for use within \addtxt{DFA} 1-RDMFT and elucidate fundamental trends in the response of different XC functionals to strong correlation in both \addtxt{DFA} 1-RDMFT and UKS-DFT.  \\

\end{abstract}

\maketitle

\section{Introduction}

Kohn-Sham Density Functional Theory (KS-DFT) has become one of the most widely available and utilized methods in computational chemistry over the past 30 years\cite{KSDFT,HK,Dumaz2021,citationstats}. This widespread adoption is due to its low computational scaling of $\mathcal{O}(N^3)$, enabling it to treat systems that are intractable for wavefunction-based methods, in combination with the relatively high accuracy of modern exchange correlation (XC) functionals.~\cite{scaling1,scaling2,ReactionBarriers,vibrations,tddft,ionization,Geometries} However, the exact XC functional is unknown and therefore approximations are used in its place. This results in significant variation in the accuracy of DFT calculations, depending on the underlying XC functional utilized and the system being considered.\cite{Mardirossian2017} Examples of common issues with which XC functionals are known to struggle include the treatment of charge transfers and strongly correlated systems, as well as a notable electron self-interaction error.\cite{failures,SIE,yang} \\

Due to the aforementioned errors and the near-ubiquity of DFT, a large number of XC functionals have been constructed over the years. These are broadly categorized using a Jacob's ladder style classification scheme, introduced by Perdew in 2001.\cite{Jacobs_ladder} Jacob's ladder classifies XC functionals according to the amount of information included in the functional. As one ascends the rungs, functionals, in theory, become more accurate. In ascending order, the first three rungs of Jacob's ladder are Local Spin Density Approximations (LSDAs), which rely only on the electron density, Generalized Gradient Approximations (GGAs), which include the gradient of the electron density, and Meta-Generalized Gradient Approximations (mGGAs), which include the kinetic energy density. Further rungs are presented by Hybrid and Range Separated Hybrid (RSH) functionals, which include exact Hartree-Fock (HF) exchange, and Double Hybrids, which introduce a wavefunction-based correlation energy correction, usually in the form of a perturbation method.\cite{Perturbation,Andreas1994} \\

The construction of functionals on any rung of Jacob's ladder generally falls into two different schools of thought.\cite{Su2017} The most common approach is to optimize the functional parameters using chemical systems to reproduce known accurate experimental or theoretical values, yielding empirical functionals. A significant difficulty with this approach is collecting a large and varied swath of accurate reference data to optimize the functional, as only using a limited number of systems and properties may result in a functional with limited applicability outside of the properties optimized on. Commonly used examples of functionals developed in this way include the Minnesota class of functionals, e.g., M06-L, M11-L, MN12-L, MN15-L.\cite{MN15L,MN06L,MN11L,MN12L} \addtxt{It has been noted that modern functional development with its focus on reproducing chemical properties has resulted in a degradation of the fundamental quantity of DFT, the electron density, suggesting that some functional improvements are the result of overfitting and do not present better approximations to the exact functional.\cite{medvedev,Hammes-Schiffer2017}} The other approach to functional construction is to accurately reproduce and obey known physical properties and conditions that the unknown exact functional must satisfy, yielding non-empirical or ab-initio functionals. Examples of functionals constructed in this way that are commonly utilized include TPSS, revTPSS, and SCAN.\cite{TPSS,revTPSS,SCAN}\\

A significant challenge for DFT is the capture of static (also referred to as strong or multi-reference correlation), which is characterized by multiple Slater determinants contributing significantly to the wavefunction. These effects are difficult to capture via improvements to the functional as they are non-local in nature.\cite{Becke2013} Due to this difficulty and the high computational scaling of systematically improvable wavefunction-based alternatives, significant efforts have been made to develop methodologies that can capture strong correlation effects while still utilizing existing XC functionals. These include the breaking of spin symmetry, allowing the alpha and beta electrons to occupy different spatial orbitals (spin-unrestricted or broken-symmetry DFT (UKS-DFT)),\cite{Perdew2021,Grafenstein2002,UHF,UHF2} enforcing the \addtxt{Perdew–Parr–Levy–Balduz (PPLB)} flat-plane conditions with fractional spins and charges \addtxt{to recover the piece wise linearity between integer numbers of electrons},\cite{Su2018,Mahler2024,PPLB} as well as utilizing complex and hyper-complex numbers to create fractionally occupied orbitals \addtxt{while still maintaining a single slater determinant reference.}\cite{Su2022,Su2021,Su2023,MHG} \addtxt{The latter two approaches both utilize fractional orbital occupations in capturing the strong correlation effects.}\\

An alternative to DFT or the more expensive wavefunction based approaches for describing electron correlation is provided by 1-electron Reduced Density Matrix (1-RDM) Functional Theory (1-RDMFT). This class of methods aims to capture strong electron correlation through fractional occupations of the 1-RDM.\cite{Gilbert, Levy1979, Valone1980, Muller1984, Goedecker1998, Mazziotti2000, Piris2007, Sharma2008, Rohr2008, Piris2017,Schmidt_2019,Buchholz_2019,RDMFT, Piris2021,Schilling_2019} \addtxt{1-RDMFTs have previously shown success in describing band gaps of semiconductors and Mott insulators, both of which DFT struggles with.\cite{Sharma2013,Pernal2015}} Natural Orbital Functional Theory (NOFT) is a related approach, in which natural orbitals and their occupation numbers are used in place of the 1-RDM to reconstruct the 2-RDM subject to N-representability conditions.\cite{PNOF1,PNOF2,PNOF3,Piris2017,Piris2021} \addtxt{Unfortunately, both of these approaches require expensive orbital optimizations under orthogonality constraints limiting the sizes of systems they can be applied to.\cite{Pernal2016} However, since these approaches use the 1-RDM as their fundamental variables, they are naturally suited to describing strong correlation through fractional occupation numbers.}\\

In a similar vein, we have previously developed a \addtxt{DFA} 1-RDMFT that combines the strengths of 1-RDMFTs to capture strong correlation through the 1-RDM with standard XC functionals, which are well suited for capturing dynamical correlation.\cite{Gibney2022,Gibney2023,Gibney2024,Gibney2022_a} This formulation allows us to \addtxt{inherit} the favorable computational scaling of \addtxt{the utilized DFA}, as well as its accuracy in non-strongly-correlated systems, \addtxt{through recovering DFT's energy,} while significantly improving upon \addtxt{it} in the presence of strong correlation. Additionally, as strong correlation is recovered through the 1-RDM, no unphysical breaking of spin-symmetry, as is the case in UKS, occurs.\\

As currently available XC functionals are not designed with \addtxt{DFA} 1-RDMFT in mind, our goal is to work toward the creation of a task-specific functional designed to perform well within the \addtxt{DFA} 1-RDMFT framework. Here, the aim is to have the XC functional recover dynamical correlation while the 1-RDM correction captures strong correlation. To this end, we expand upon previous work,\cite{Gibney2022_a,Gibney2023,Gibney2024} \addtxt{which focused on developing the DFA 1-RDMFT methodology and its derivation} \addtxt{through} exploration of the XC functional dependence in the \addtxt{DFA} 1-RDMFT framework by benchmarking nearly 200 XC functionals. This large scope, spanning the rungs of Jacob's ladder, allows us to elucidate trends in functional response to multi-reference character and identify general scaling parameters, $\kappa$ (\textit{vide infra}), for a large class of functionals commonly used today. Here, the value of $\kappa$, correlates with the magnitude of the correction required for a specific functional to recover strong correlation effects through the 1-RDM term. We compare the functional errors obtained using both UKS-DFT and \addtxt{DFA} 1-RDMFT to uncover any functional relationships between the two. Finally, the relationship between \addtxt{DFA} 1-RDMFT and HF exchange is explored through systematic modification of hybrid XC functionals.\\

\section{Theory}
The \addtxt{DFA} 1-RDMFT energy is defined as
\begin{equation}
    E_{1RDMFT}[{}^1D] = E_{DFT}[{}^1D] + \tilde{w}\text{Tr}({}^1D^2-{}^1D)\,.
    \label{eq:energy}
\end{equation}
The first term, $E_{DFT}[{}^1D]$, is the traditional DFT energy expressible as:
\begin{equation}
    E_{DFT}[{}^1D] = T_s[{}^1D] + V[\rho] + F_{xc}[\rho]\,,
\end{equation}
with $T_s$ being the non-interacting kinetic energy, $V[\rho]$ being the energy from the electron-electron, electron-nuclei, and nuclei-nuclei columbic energies, and $F_{XC}[\rho]$ being the XC energy. The second term, $\tilde{w}\text{Tr}({}^1D^2-{}^1D)$, is the 1-RDM functional, which we derived previously from the first term in the unitary decomposition of the 2-electron reduced density matrix's cumulant contribution to the electronic energy.\cite{Gibney2023,Gibney2024} \addtxt{We have shown previously that this term corresponds primarily to strongly correlated electronic interactions, due to it only including the diagonal terms in the 2-RDMs cumulant.\cite{Gibney2023,mazziotti2008}} $\tilde{w}$ above is defined as
\begin{equation}
\label{eq:cg2}
{\tilde w} = \frac{{\kappa\rm Tr}(^{2} I)}{{\rm Tr}(^{2} W \, ^{2} I)} \sum_{ij}(2\langle ij|ij\rangle - \langle ij|ji\rangle)\,,
\end{equation}
where
\begin{equation}
{\rm Tr}(^{2} W \, ^{2} I) = 4 \sum_{\addtxt{\tilde{i}}<\addtxt{\tilde{j}}}{{\tilde W}_{\addtxt{\tilde{i}\tilde{j}}}} + \sum_{\addtxt{\tilde{i}}}{{\tilde W}_{\addtxt{\tilde{i}\tilde{i}}}}\,,
\end{equation}
\begin{equation}
{\tilde W}_{\addtxt{\tilde{i}\addtxt{j}}} = \frac{\langle \addtxt{\tilde{i}\tilde{j}}||\addtxt{\tilde{i}\tilde{j}}\rangle}{\sqrt{\langle \addtxt{\tilde{i}\tilde{i}}||\addtxt{\tilde{i}\tilde{i}}\rangle \langle \addtxt{\tilde{j}\tilde{j}}||\addtxt{\tilde{j}\tilde{j}}\rangle}}\,,
\end{equation}
and
\begin{equation}
{\rm Tr}(^{2} I) = \frac{r(r-1)}{2}.
\end{equation}
Here, $r$ denotes the number of spin orbitals, $i$ and $j$ run over the spatial orbitals, \addtxt{$\tilde{i}$ and $\tilde{j}$ run over the atomic orbitals,} and $\kappa$ is a functional specific scaling parameter between 0 and 1.
The overall 1-RDMFT energy inherits DFTs non-linearity in ${}^1D$ and it must therefore be obtained via a self-consistent field procedure with the objective function minimized at each iteration being: 
\begin{equation}
    \text{min}~\text{Tr}({}^1H_{KS}[{}^1D]{}^1D) + \tilde{w}\text{Tr}({}^1D^2-{}^1D)\,,
    \label{eq:w_tilde}
\end{equation}
where ${}^1H_{KS}$ is the Kohn-Sham Hamiltonian. This yields a new ${}^1D$, which is used to construct the next ${}^1H_{KS}[{}^1D$]. This process is repeated until convergence. 
\addtxt{It should be noted that $\tilde{w}$ only needs to be calculated once, and scales as $\mathcal{O}(N^4)$ due to its dependence on the 2-electron integrals. In practice, the calculation of $\tilde{w}$ is fast, with the main bottleneck for the DFA 1-RDMFT procedure being the iterative solving of Equation \ref{eq:w_tilde}.} We implement the above minimization for ${}^1D$ as a linear Semi-Definite Program (SDP), applying the appropriate N-representability constraints such that the resulting ${}^1D$ is N-representable\cite{NRep}.\\

As can be seen from the quadratic form of the 1-RDM functional, the 1-RDMFT correction to the energy is only non-zero when idempotentency in the 1-RDM (${}^1D^2={}^1D$) is broken. This allows the \addtxt{DFA} 1-RDMFT to reproduce traditional KS-DFT results in the absence of strong correlation, where the 1-RDM is idempotent. Consequently, the \addtxt{DFA} 1-RDMFT framework retains the accuracy and scaling of traditional KS XC functionals in weakly correlated systems, while compensating static correlation errors in the presence of multi-reference correlation.\\

\section{Computational Methodology}
All \addtxt{DFA} 1-RDMFT calculations were performed \addtxt{in a spin restricted framework} using in-house developed code, which utilizes PySCF\cite{PySCF} for the integral generation, LibXC\cite{LibXC} to evaluate the XC functionals, and CVXPY\cite{diamond2016cvxpy} with the SCS\cite{scs} solver for the SDP.\cite{ocpb:16} \addtxt{An example comparing the timings of DFA 1-RDMFT against UKS DFT, both utilizing the PBE functional from PySCF, along with a discussion of the results, are available in the SI as Table S3 and Figure S4.} UKS DFT calculations were performed using ORCA 6.0.0\cite{orca} in combination with LibXC. Reference results were obtained with the Anti-Hermitian Contracted Schrodinger Equation (ACSE).\cite{ACSE,ACSE1,ACSE2,ACSE3} The cc-pVDZ basis set was utilized for all calculations.\cite{ccpvdz} \\

\section{Results}
In this work, we consider the dissociations of C$_2$, N$_2$, CN, F$_2$, NO, S$_2$, SiO, and CO. The set of investigated functionals presents a subset of those available in the LibXC library. It includes only full XC functionals, which were denoted with "\_XC\_" in their name or separate exchange "\_X\_" and correlation "\_C\_" components with identical names, which were combined to create the full XC functional. This yields 217 distinct functionals. After removing functionals that failed to converge for any point in any of the systems for 1-RDMFT or DFT, this yields a final set comprising 190 functionals. The functionals used can be broken down into 6 LDAs, 50 GGAs, 15 mGGAs, 78 Hybrids, and 41 RSHs. To ensure that we are comparing differences in the functionals themselves, we equalize the effect of the 1-RDM term across the test set by optimizing the scalar $\tilde{w}$ value in \addtxt{DFA} 1-RDMFT for each system/functional combination individually to reproduce the dissociation energy as obtained from the ACSE. The optimized $\tilde{w}$ value is then held fixed throughout all internuclear distances for each system/functional combination to generate the dissociation curves from equilibrium to 3 \AA~for both \addtxt{DFA} 1-RDMFT and UKS-DFT. \addtxt{This ensures that we obtain the optimal performance for each XC functional in DFA 1-RDMFT, enabling us to identify trends across Jacob’s ladder and $\kappa$.} \addtxt{Importantly, this enables us to explore the interplay between the DFA utilized and the 1-RDM corrective term to infer how well correlation is being captured along the entirety of the dissociation curve.}\\

\begin{figure}
    \centering
    \includegraphics[width=\linewidth]{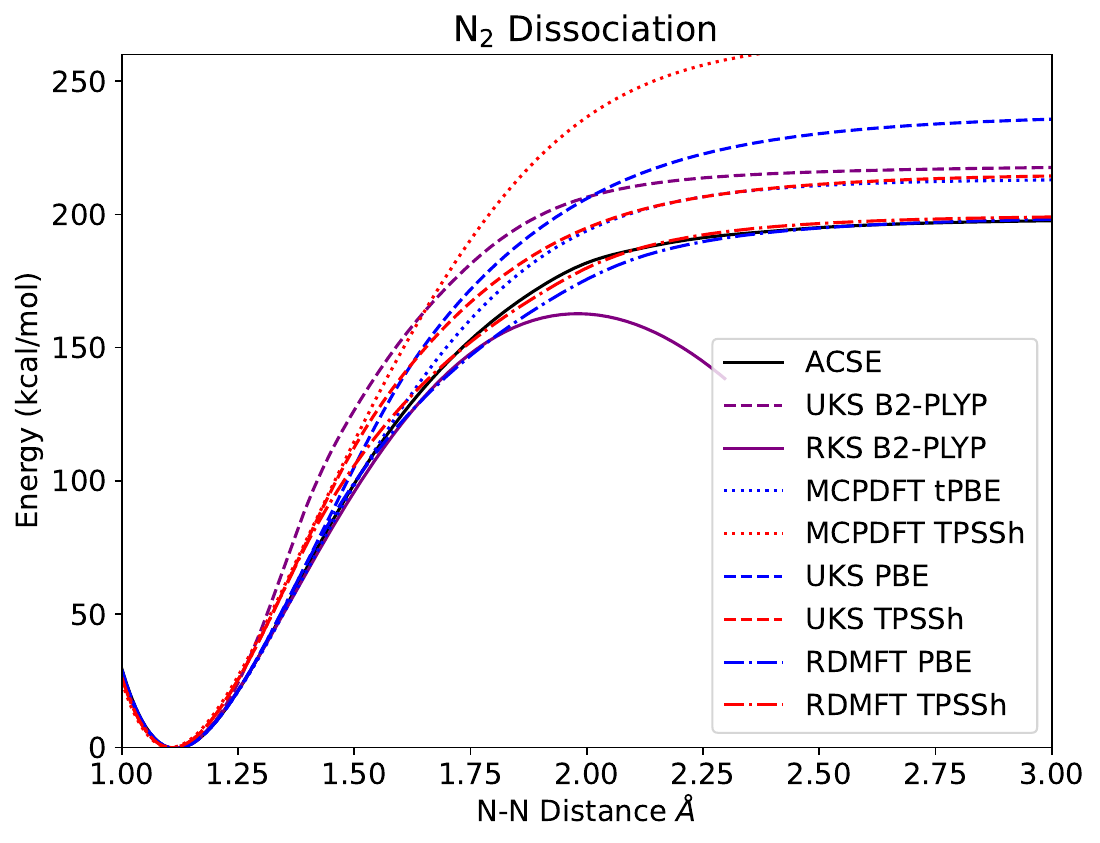}
    \caption{\addtxt{N$_2$ dissociation curves comparing the DFA 1-RDMFT, UKS DFT, MCPDFT, and ACSE.}}
    \label{fig:dissociation}
\end{figure}

\addtxt{An example of these dissociations (N$_2$) is presented in Figure \ref{fig:dissociation}. To place this work within the broader realm of available techniques, Figure \ref{fig:dissociation} also includes the double hybrid B2-PLYP\cite{Grimme2006,Grimme2007} in both the spin restricted and unrestricted formalisms, as well as MC-PDFT with the translated tPBE and tTPSSh functionals\cite{MCPDFT1,MCPDFT2}. B2-PLYP, in the restricted KS framework gives excellent performance compared to the reference ACSE curve around equilibrium but breaks down as the bond is stretched. B2-PLYP in the unrestricted KS framework yields a better dissociation limit but displays significant deviations from the ACSE curve in the stretched bond region. MC-PDFT tPBE improves upon UKS PBE, however, MC-PDFT tTPSSh yields larger errors than UKS TPSSh. UKS DFT with both PBE and TPSSh overestimates the dissociated limit. Around the equilibrium geometry, UKS PBE recovers the ACSE potential energy surface whereas UKS TPSSh shows deviations from the ACSE throughout its entirety. Transitioning to DFA 1-RDMFT not only reproduces the dissociated limit, which is the target utilized to determine $\kappa$, but also improves upon UKS DFT in the stretched bond region. In the absence of multireference effects (up to R(N-N) = 1.4 \AA~), DFA 1-RDMFT recovers UKS DFT as the 1-RDM remains idempotent.}

Zeroing all curves using their equilibrium values to facilitate comparisons between functionals and the ACSE, we calculated two error metrics: the maximal error between either the \addtxt{DFA} 1-RDMFT or DFT curve and the reference ACSE curve
\begin{equation}
    M_e = max |E_{ACSE}(r) - E_{RDMFT}(r)|\,,
\end{equation} 
and the cumulative error between the two curves
\begin{equation}
    C_a = \int_{eq}^{3 \AA}|E_{ACSE}(r) - E_{RDMFT}(r)|dr\,.
\end{equation}

\begin{figure}
    \centering
    \includegraphics[width=\linewidth]{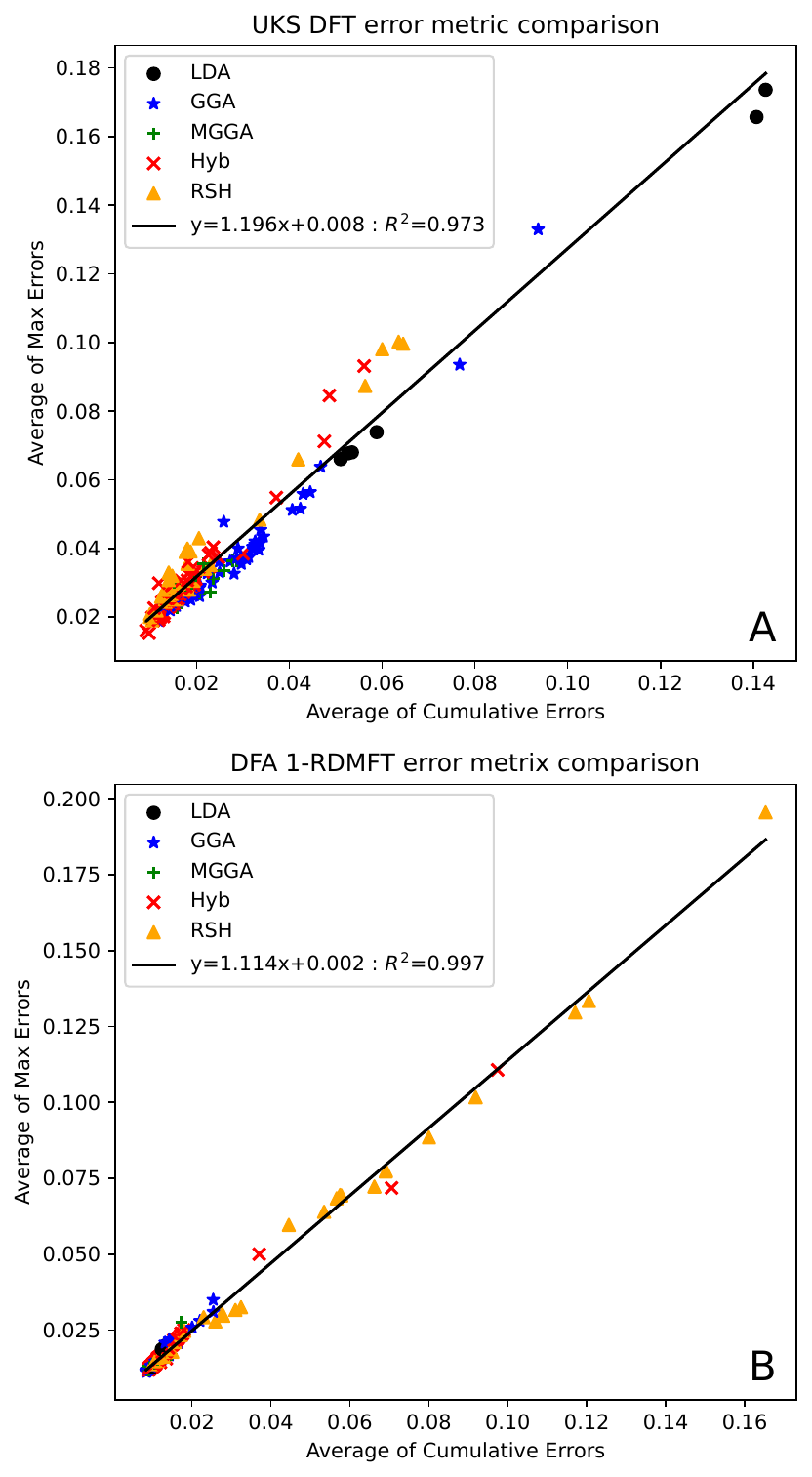}
    \caption{Plots of the average of max errors vs the average of cumulative errors, in Hartree, for (A): UKS-DFT, and (B): \addtxt{DFA} 1-RDMFT results for all systems and functionals used in this work.}
    \label{fig:error_metrics}
\end{figure}

To quantify the maximal error and cumulative error metrics, we compare their averages across all systems for each functional in Figure \ref{fig:error_metrics}. We observe a nearly linear relationship between the two error metrics, indicating that either is equally valid for the remainder of this work. This also implies that the errors are not localized to a single point along the dissociation but are distributed along it. Plotting the linear regression and $R^2$ values, we find that, while both show $R^2s > 0.95$, \addtxt{DFA} 1-RDMFT is more linear with a $R^2$ of 0.997 vs. UKS-DFT's 0.973. This is likely due to \addtxt{DFA} 1-RDMFT being able to capture the dissociated limit for all systems whereas DFT, even with spin-symmetry breaking, is unable to achieve this for all systems. Additionally, the slope for UKS-DFT's errors is greater than \addtxt{DFA} 1-RDMFT's, 1.196 vs 1.114, indicating the max error is increasing faster than the cumulative error. Again, this is likely due to UKS-DFT not capturing the dissociated limit in all systems. For the remainder of this work, we will utilize the cumulative error as our key metric in the text; max error data may be found in the supplemental information (SI).\\

\begin{figure}
    \centering
    \includegraphics[width=\linewidth]{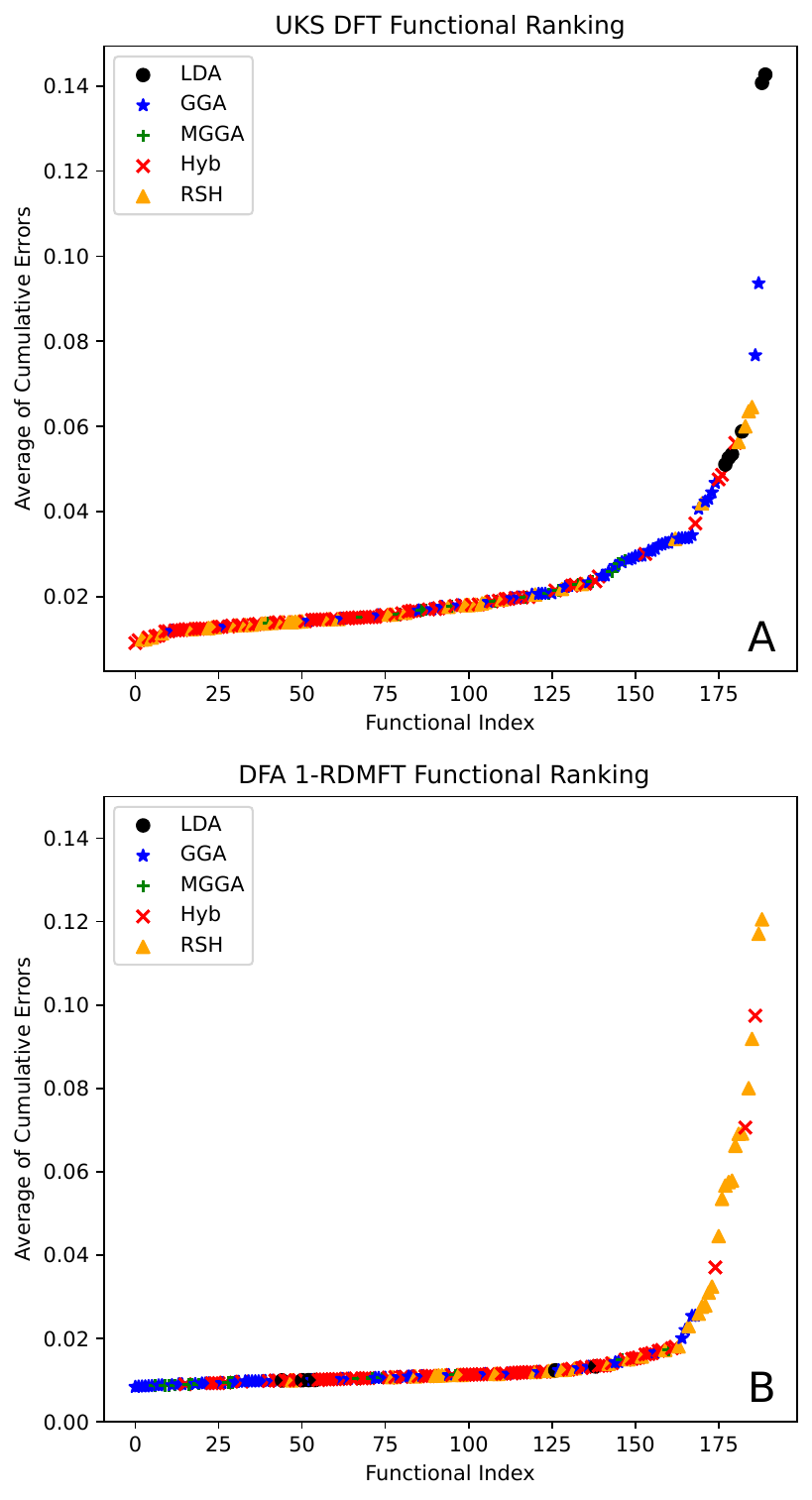}
    \caption{Plots the functional rankings for (A): UKS-DFT, and (B): \addtxt{DFA} 1-RDMFT. The functionals are ranked in order of increasing average of cumulative errors in Hartree.}
    \label{fig:functionals_ranked}
\end{figure}

In Figure \ref{fig:functionals_ranked} we sort the functionals according to their cumulative error and classify them according to their placement on Jacob's ladder. \addtxt{The functionals ranks are tabulated and available in Table S2 of the SI}. Starting with UKS-DFT, we notice that the best performing functionals are primarily of Hybrid and RSH nature, with mGGA functionals being interspersed throughout. The worst performing functionals are dominated by the GGA and LDA functionals. In contrast, within \addtxt{DFA} 1-RDMFT, the functional ordering is largely reversed, with the worst performing functionals comprising mainly Hybrid and RSH functionals, while GGAs and MGGAs yield the best performance. LDA based functionals are interspersed throughout. The cumulative errors for UKS-DFT show a greater slope than those for \addtxt{DFA} 1-RDMFT, indicating a greater XC functional dependence in UKS-DFT compared to \addtxt{DFA} 1-RDMFT, where this dependence is dampened. Analysis of the max errors yields identical conclusions and is presented in the SI (Figure S1). \\

\begin{figure}
    \centering
    \includegraphics[width=\linewidth]{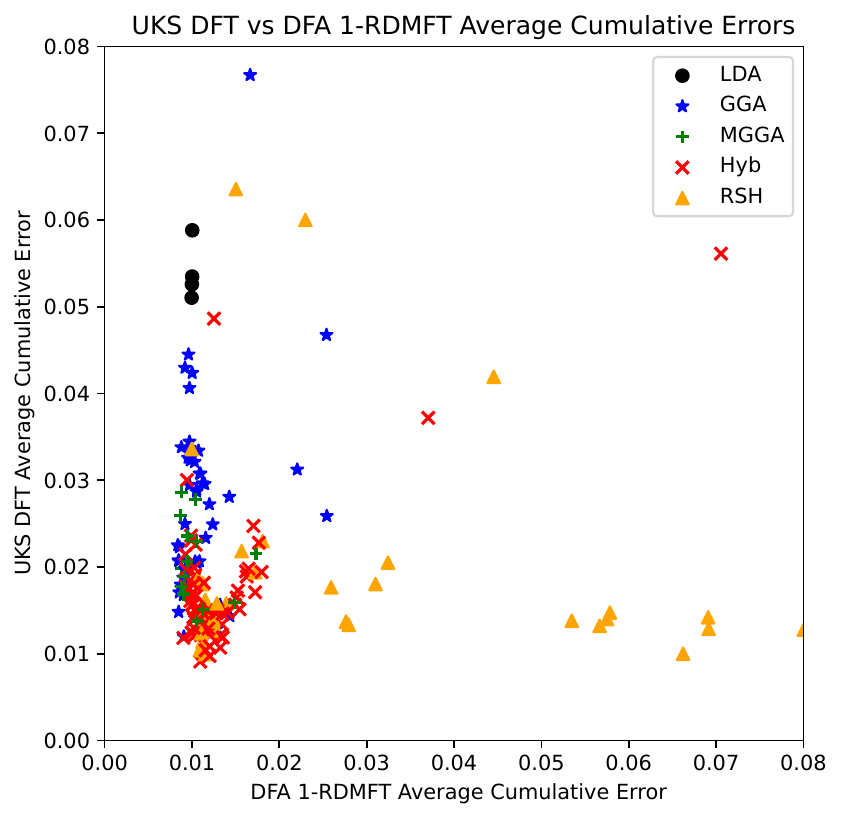}
    \caption{Comparison of the average cumulative errors, in Hartree, from UKS-DFT and \addtxt{DFA} 1-RDMFT for the majority of functionals investigated in this work. Three and four functionals from UKS-DFT and \addtxt{DFA} 1-RDMFT, respectively, with errors $>$ 0.08, are omitted from the figure to improve readability.}
    \label{fig:UKSvsRDMFT}
\end{figure}

Comparison of cumulative functional errors between UKS-DFT and \addtxt{DFA} 1-RDMFT, as shown in Figure \ref{fig:UKSvsRDMFT}, reveals that there is no significant correlation between them. \addtxt{Local functionals span a wide range of errors in UKS DFT but that spread decreases notably when moving from LDAs to GGAs and ultimately mGGAs. This is further emphasized by the standard deviations of UKS DFT errors, which are presented in Table \ref{tab:avg_rung}, and are 0.0454, 0.0144, and 0.0046 for LDAs, GGAs, and mGGAs, respectively. Although the local functionals display a wide range of errors in UKS DFT, in DFA 1-RDMFT their error spread is reduced with standard deviations of 0.0015, 0.0040, and 0.0025 for LDAs, GGAs, and mGGAs, respectively. Including exact exchange with hybrid functionals yields standard deviations of 0.0081 and 0.0121 for UKS DFT and DFA 1-RDMFT, respectively, presenting a departure from the behavior of pure functionals, where DFA 1-RDMFT outperforms UKS-DFT. The effects of exact exchange on the DFA 1-RDMFT framework is explored in more detail in the later part of this study.} \addtxt{These results} indicate that performance of an XC functional in UKS-DFT holds \addtxt{limited} predictive power for its performance in \addtxt{DFA} 1-RDMFT. Importantly, this emphasizes that UKS-DFT and \addtxt{DFA} 1-RDMFT are fundamentally different in the ways they capture correlation, through the XC functional and the breaking of spin-symmetry in UKS-DFT, and through fractional occupations in the 1-RDM and an electron integral derived term in \addtxt{DFA} 1-RDMFT. The max errors, available in the SI (Figure S2), again display an identical trend.\\

Using the Jacob's ladder classification scheme, we compare the functionals with the lowest cumulative error per rung in Table \ref{tab:Best_funcs} \addtxt{along with the \addtxt{DFA} 1-RDMFT optimized $\kappa$ values}. LDA presents the only rung for which the lowest errors for both UKS-DFT and \addtxt{DFA} 1-RDMFT are achieved with the same functional (Teter93). This is likely due to the limited number of LDAs available (six) in comparison to other rungs. Here, a direct comparison shows that the Teter93 functional performs significantly better when used within 1-RDMFT compared to UKS-DFT. Indeed, comparing the best performing functional on each rung between UKS-DFT and \addtxt{DFA} 1-RDMFT reveals that \addtxt{DFA} 1-RDMFT always displays a lower cumulative error except for when using a RSH functional, where they display identical errors. The range of errors within \addtxt{DFA} 1-RDMFT is notably lower than in UKS-DFT at 0.0016 Hartree vs 0.0419 Hartree, respectively. Interestingly, while UKS-DFT tends to decrease in cumulative error as one ascends Jacob's ladder, the opposite is found for \addtxt{DFA} 1-RDMFT. \addtxt{Inspection of the performance of commonly utilized, popular functionals reveals TPSSh and N12 as good candidates for common use cases, where, for example, geometries are optimized at the DFT level while further electronic properties are obtained from subsequent DFA 1-RDMFT single point calculations. It should be noted that DFA 1-RDMFT includes the UKS DFT error around equilibrium as in weakly correlated regions where no breaking of spin symmetry occurs in UKS DFT DFA 1-RDMFT reproduces the DFT energies.} \\

\begin{table}[]
    \centering
\begin{tabular}{l|c|c|c|c|c}
         & \multicolumn{2}{c}{UKS DFT} & \multicolumn{3}{c}{\addtxt{DFA} 1-RDMFT} \\\hline
    Rung & Func       & Error & Func      & Error    & \addtxt{$\kappa$} \\\hline
    LDA  & teter93    & 0.0510 & teter93   & 0.0100  & \addtxt{0.1118} \\
    GGA  & n12        & 0.0120 & hcth\_p14 & 0.0084  & \addtxt{0.1048} \\
    mGGA & mn15       & 0.0138 & tm        & 0.0087  & \addtxt{0.1029}\\
    Hyb  & b1lyp      & 0.0091 & tpssh     & 0.0090  & \addtxt{0.1347} \\
    RSH  & camh-b3lyp & 0.0100 & tuned-cam-b3lyp & 0.0100 & \addtxt{0.2081}     \\
    \end{tabular}
    \caption{Functionals with the lowest cumulative error, in Hartree, per rung of Jacob's ladder as obtained with UKS DFT and \addtxt{DFA} 1-RDMFT and their associated cumulative error. \addtxt{DFA 1-RDMFTs optimal $\kappa$ values are also included.}}
    \label{tab:Best_funcs}
\end{table}

To further explore the \addtxt{effect of the XC functional form} on the accuracy of UKS-DFT and \addtxt{DFA} 1-RDMFT calculations, we investigate the average cumulative errors and their standard deviations per rung of Jacob's ladder (Table \ref{tab:avg_rung}). This analysis reveals that \addtxt{DFA} 1-RDMFT's errors generally are significantly lower than those obtained from UKS-DFT. Again, the only rung that presents an exception to this trend is the RSHs, where 1-RDMFT performs worse than UKS-DFT. Critically, within \addtxt{DFA} 1-RDMFT, no large differences between the local functionals (LDA, GGA, and MGGA) are observed, but errors increase significantly when introducing hybrid and especially RSH functionals. In contrast, the best performing rung for UKS-DFT is hybrid functionals. Comparison of standard deviations identifies that \addtxt{DFA} 1-RDMFT significantly reduces the differences in the errors for the local functionals with respect to UKS-DFT, while increasing the deviations for the Hybrid and RSH functionals. \addtxt{We believe the improvement in the performance of local functionals with DFA 1-RDMFT may primarily be due to a more accurate representation of the underlying electron density introduced by fractional orbital occupations, as was argued in previous work that allowed fractional occupations without the 1-RDM energetic correction\cite{gibney2020}. The increase in errors in hybrid functionals is thought to derive from the 1-RDM corrective term’s dependence on HF exchange and the use of a spin-restricted 1-RDM (\textit{vide infra}).} \\

\begin{table}[]
    \centering
\begin{tabular}{l|c|c|c|c}
         & \multicolumn{2}{c}{UKS DFT} & \multicolumn{2}{c}{\addtxt{DFA} 1-RDMFT}\\\hline
    Rung & Error   & Stdev& Error  & Stdev\\\hline
    LDA  & 0.0832  &0.0454& 0.0109 &0.0015\\
    GGA  & 0.0291  &0.0144& 0.0113 &0.0040\\
    mGGA & 0.0204  &0.0046& 0.0104 &0.0025\\
    Hyb  & 0.0173  &0.0081& 0.0140 &0.0121\\
    RSH  & 0.0207  &0.0147& 0.0364 &0.0364\\
    \end{tabular}
    \caption{The average cumulative errors and standard deviations, in Hartree, for each rung of Jacob's ladder for both UKS DFT and \addtxt{DFA} 1-RDMFT.}
    \label{tab:avg_rung}
\end{table}

\begin{figure}
    \centering
    \includegraphics[width=\linewidth]{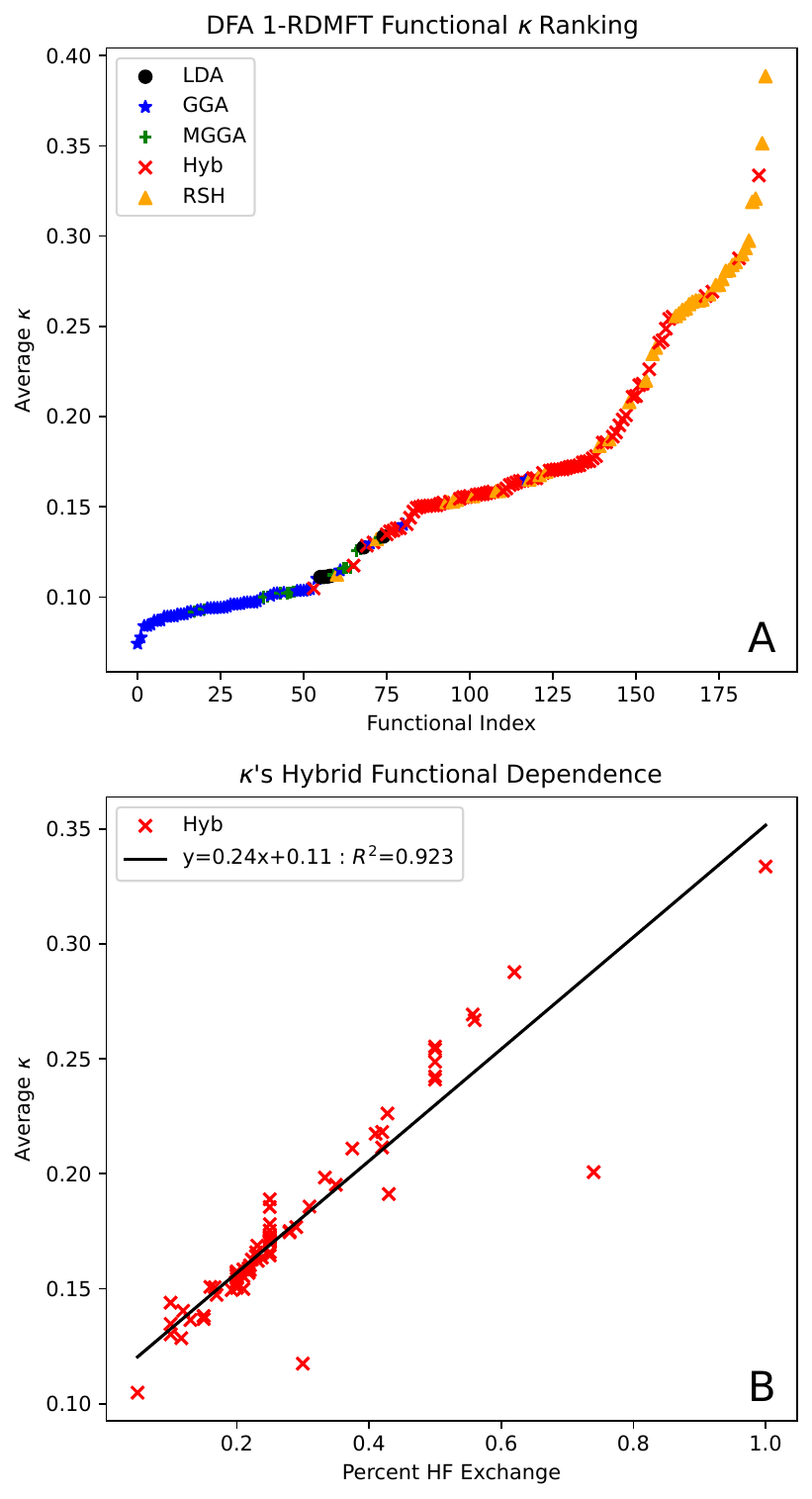}
    \caption{(A): Plot of the functional ranking according to the optimal average $\kappa$ value, and (B): a plot of the relationship between HF exchange and the $\kappa$ value.}
    \label{fig:alphas}
\end{figure}

\addtxt{Since the DFA 1-RDMFT energy is dependent on both the XC and 1-RDM functionals, the varying $\kappa$ values indicate the balance struck between correlation effects captured via the XC functional and those captured by the 1-RDM functional, with a larger value of $\kappa$ denoting a stronger 1-RDM correction.} As $\tilde{w} = \frac{{\kappa\rm Tr}(^{2} I)}{{\rm Tr}(^{2} W \, ^{2} I)} \sum_{\addtxt{\tilde{i}\tilde{j}}}(2\langle \addtxt{\tilde{i}\tilde{j}}|\addtxt{\tilde{i}\tilde{j}}\rangle - \langle \addtxt{\tilde{i}\tilde{j}}|\addtxt{\tilde{j}\tilde{i}}\rangle)$ is derived from the electron repulsion integrals, and was optimized for each system/functional combination separately, we can use this to calculate the optimal $\kappa$ value for each functional. The results of this analysis are plotted in Figure \ref{fig:alphas}. We again find that functionals of the same rung tend to cluster together with similar $\kappa$ values. As is evident from the figure, GGAs and mGGAs require the smallest $\kappa$ values and thus the least correction from the 1-RDM term, whereas hybrid and RSH functionals both require significantly stronger corrections.\\

\begin{figure}
    \centering
    \includegraphics[width=\linewidth]{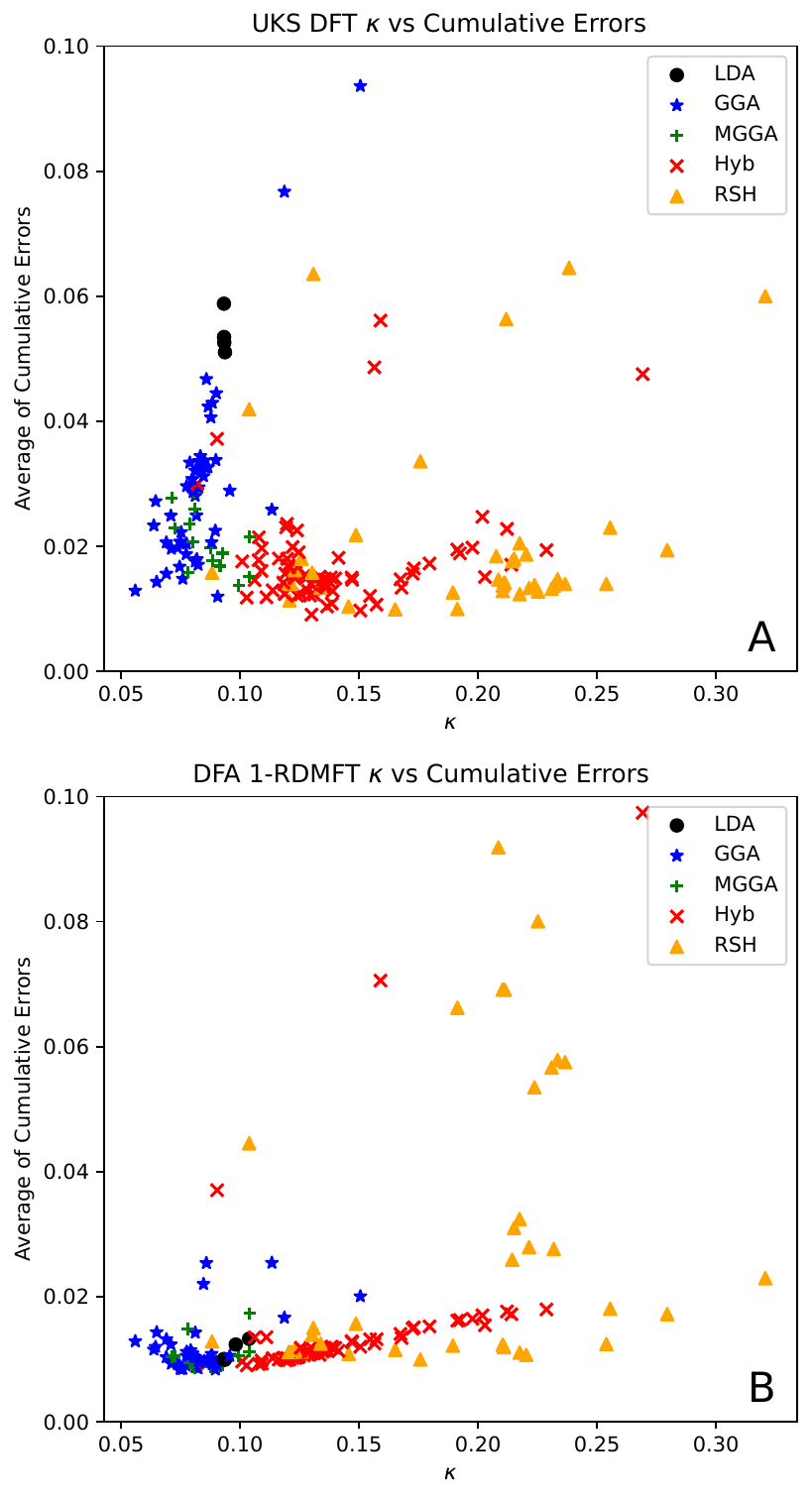}
    \caption{Plots of the cumulative error in Hartree vs $\kappa$ for (A): UKS-DFT and (B): \addtxt{DFA} 1-RDMFT. Two and three functionals with errors $>$ 0.1 in UKS-DFT and \addtxt{DFA} 1-RDMFT, respectively, are omitted from the figure to improve readability.}
    \label{fig:k_vs_Error}
\end{figure}

To determine whether there is a relationship between the optimal $\kappa$ values and the cumulative errors, we plot both for \addtxt{DFA} 1-RDMFT and UKS-DFT in Figure \ref{fig:k_vs_Error}. We include UKS-DFT here as, although UKS-DFT does not have a dependence on $\kappa$, it may correlate with the XC functional's inherent ability to capture strong correlation in the spin-symmetry broken regime. We find that in UKS-DFT, GGAs, while constrained to only a small range of optimal $\kappa$ values, display high variability in the corresponding error. Hybrids, on the contrary, span a wider range of $\kappa$ values but show a tighter grouping in terms of error. As there is no clear trend between the $\kappa$ value and the resulting cumulative error, this implies that $\kappa$ does not relate to an XC functional's ability to capture strong correlation in UKS-DFT. In \addtxt{DFA} 1-RDMFT, $\kappa$ values for LDAs, GGAs, and MGGAs are closely clustered together in the 0.05 to 0.10 range. The observed minimum in the errors suggests that a $\kappa$ value of approximately 0.075 yields the lowest errors with traditional XC functionals in \addtxt{DFA} 1-RDMFT. This implies that there is an optimal value of $\kappa$ such that each functional, XC and 1-RDM, contribute, in a balanced manner, to the overall description of electron correlation. As the 1-RDM correction is zero in the weakly correlated regime, the XC functional and $\kappa$ must facilitate a smooth transition in the capture of electron correlation as the 1-RDM term becomes increasingly active to minimize the overall error. Hybrid functionals do not display a clear $\kappa$ minimum, instead showing a clearer relationship between cumulative errors and $\kappa$. Plotting the percent HF exchange for the functional vs its cumulative error (available in the SI, Figure S3) reproduces this trend, with the smallest errors occurring for functionals with HF exchange $<$ 20\%. For both UKS-DFT and \addtxt{DFA} 1-RDMFT, RSHs results show no discernible trends.\\

Plotting the Hybrid functionals according to their percent HF exchange reveals a notable trend between the optimal $\kappa$ value and the amount of HF exchange included in the XC functional (Figure \ref{fig:alphas}). A\addtxt{n} R$^2$ value of 0.923 implies that $\kappa$ has a near linear dependence on the underlying amount of HF exchange in a given functional. Equation \ref{eq:w_tilde} gives insight into why this is the case; $\tilde{w}$ is part of the corrective 1-RDM term, which has a dependence on the exchange integrals. Since ${}^1D^2-{}^1D$ is always non-positive, the 1-RDM correction acts \addtxt{against the increase in} the exchange contribution. Therefore, increasing the amount of exchange in a functional implies a larger correction, with $\kappa$ or $\tilde{w}$ increasing. \\

\begin{figure}[h]
    \centering
    \includegraphics[width=\linewidth]{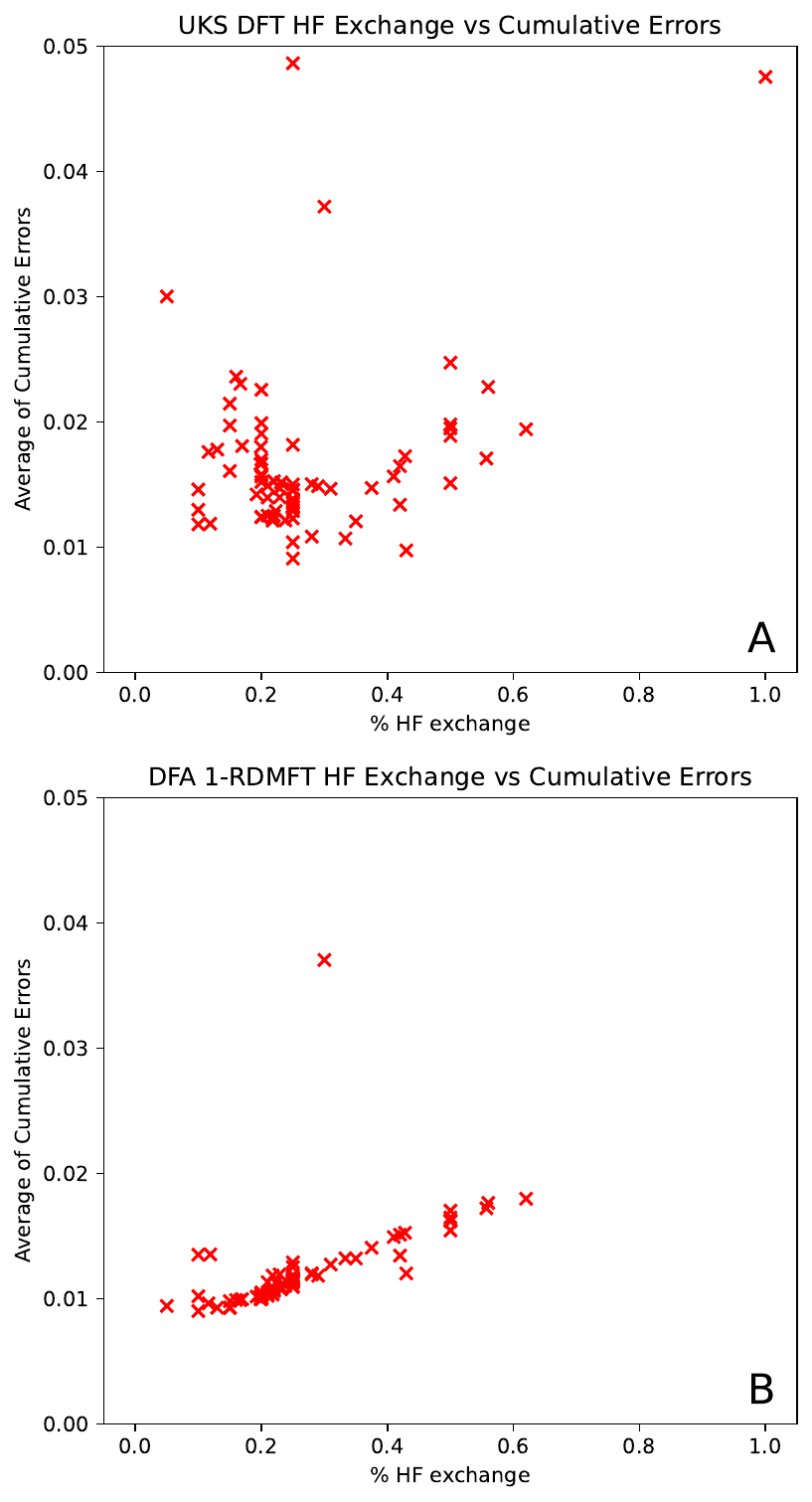}
    \caption{\addtxt{Plots of the cumulative error in Hartree vs HF exchange for (A): UKS-DFT and (B): DFA 1-RDMFT. One and two functionals with errors $>$ 0.05 in UKS-DFT and DFA 1-RDMFT, respectively, are omitted from the figure to improve readability.}}
    \label{fig:HF_error}
\end{figure}

\addtxt{To identify if this relationship extends to the errors in DFA 1-RDMFT, we plot the Hybrid functionals’ percent HF exchange against the average of cumulative errors in Figure \ref{fig:HF_error} for both UKS DFT and DFA 1-RDMFT. Although HF exchange increases static correlation errors in DFT,\cite{Slater1972, Ziegler1991, Tschinke1990} this issue can be masked by the breaking of spin symmetry.\cite{Truhlar2020} As a result, no significant dependence of the cumulative errors on the amount of HF exchange used within UKS DFT is observed. In contrast, in DFA 1-RDMFT, there is a strong relationship between the cumulative error and the amount of HF exchange. The cumulative error is minimized with $\approx$ 10\% HF exchange. Increases in the HF exchange fraction beyond this point result in larger cumulative errors until approaching an asymptotic limit. This difference in behavior from UKS DFT is likely caused by a combination of two factors: the use of a spin-restricted framework within DFA 1-RDMFT, and the underlying dependence of the 1-RDM correction term on the exchange integrals.}\\

To further elucidate the dependence on HF exchange, we selected four different XC functionals (PBE, TPSS, SCAN, and B3LYP) and varied the fraction of HF exchange according to:
\begin{equation}
    E_{x} = (1-\lambda)F_{X} + \lambda HF_{X}\,,
\end{equation}
where $F_{X}$ is the XC functional's non-HF exchange and $HF_{X}$ is HF exchange. We varied $\lambda$ from 0 to 1 in increments of 0.05 for each functional and re-optimized $\tilde{w}$, allowing the quantification of its dependence on the HF exchange fraction included in the XC functional. Multiple functionals with different base correlation and exchange components were used to identify how their differences contribute to $\kappa$.\\

\begin{figure}[h]
    \centering
    \includegraphics[width=\linewidth]{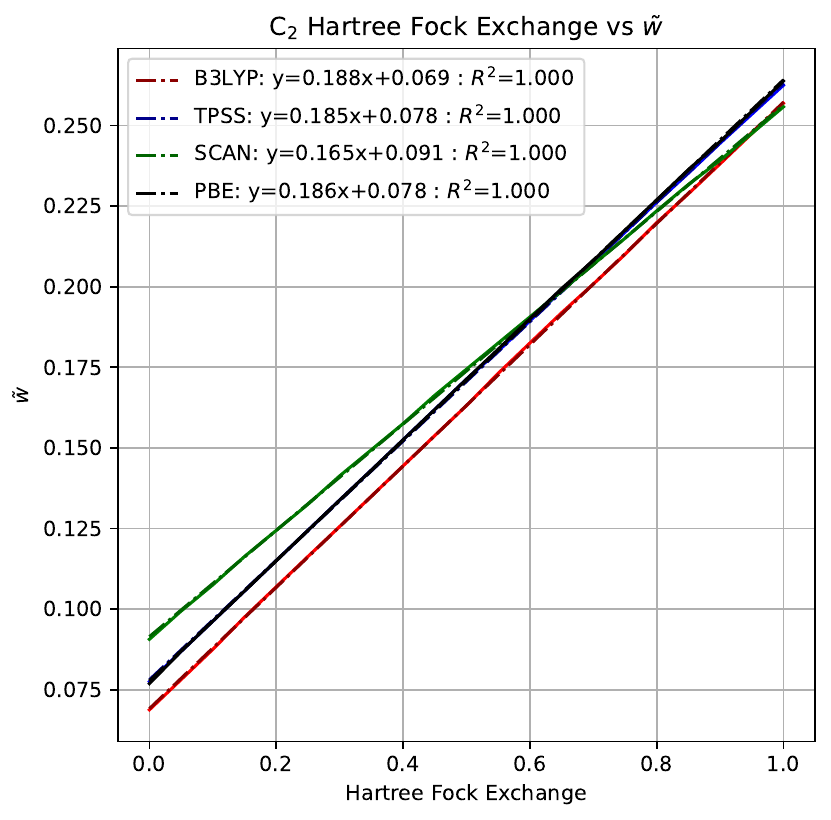}
    \caption{Plot of functional HF exchange vs optimal $\tilde{w}$ value for C$_2$. Linear regressions are also plotted and nearly perfectly cover the underlying data. TPSS and PBE data lie effectively on top of each other.}
    \label{fig:HF_exchange}
\end{figure}

Examples of the regressions obtained are displayed in Figure \ref{fig:HF_exchange}. The results shown for C$_2$ are representative of all other systems; the relationship between HF exchange and the optimal $\tilde{w}$ value is linear, with a R$^2$ value greater than 0.999 for all system. Comparing the slopes between the systems (shown in Table \ref{tab:HF_exchange}), we note that the slopes for B3LYP, TPSS, and PBE are nearly identical when applied to the same system. SCAN presents the only outlier, with its slope consistently $\approx$ 0.02 lower than the other functionals. \\

The aforementioned scans included the point $\lambda$ = 1, where each functional contains only HF exchange. At this point, the only difference between functionals arises from their correlation terms. The fact that the $\tilde{w}$ values at this point are not identical confirms that $\tilde{w}$ depends on the correlation term in the underlying XC functional. This is the result of the interplay between the XC and 1-RDM functionals. Different XC functionals \addtxt{will} capture varying fractions of the correlation to be described by the 1-RDM functional\addtxt{\cite{Gritsenko1997}}. To further investigate this, we compare the $\lambda$ = 1 point for all functionals (available in the SI (Table S1)), which reveals that different correlation terms only result in small differences in the $\tilde{w}$ value, on the order of $\approx$ 0.01. These results imply that the overlapping correlation obtained by both the XC and 1-RDM functional is relatively minor. In contrast, changing the system while utilizing the same XC functional, results in significantly larger changes to $\tilde{w}$ of up to $\approx{0.1}$.

\begin{table}[]
    \centering
    \begin{tabular}{l|c|c|c|c}
    System & \multicolumn{4}{c}{Functional}\\
         & B3LYP  &  PBE  &  SCAN & TPSS\\\hline
    C$_2$& 0.188  &	0.186 &	0.165 &	0.185\\
    CN	 & 0.220&	0.218 &	0.197 &	0.216\\
    CO	 & 0.246  &	0.246 &	0.223 &	0.242\\
    F$_2$& 0.315&	0.316 &	0.294 &	0.309\\
    N$_2$& 0.246  &	0.246 &	0.224& 0.243\\
    NO	 & 0.259&	0.259&	0.237& 0.256\\
    S$_2$&	0.159& 0.158&	0.145&	0.156\\
    SiO	 & 0.201&	0.202&	0.182 &	0.197\\
    \end{tabular}
    \caption{Slope of the linear regression between the optimal value of $\tilde{w}$ and HF exchange. All R$^2$ values are greater than 0.999.}
    \label{tab:HF_exchange}
\end{table}

\begin{figure}
    \centering
    \includegraphics[width=\linewidth]{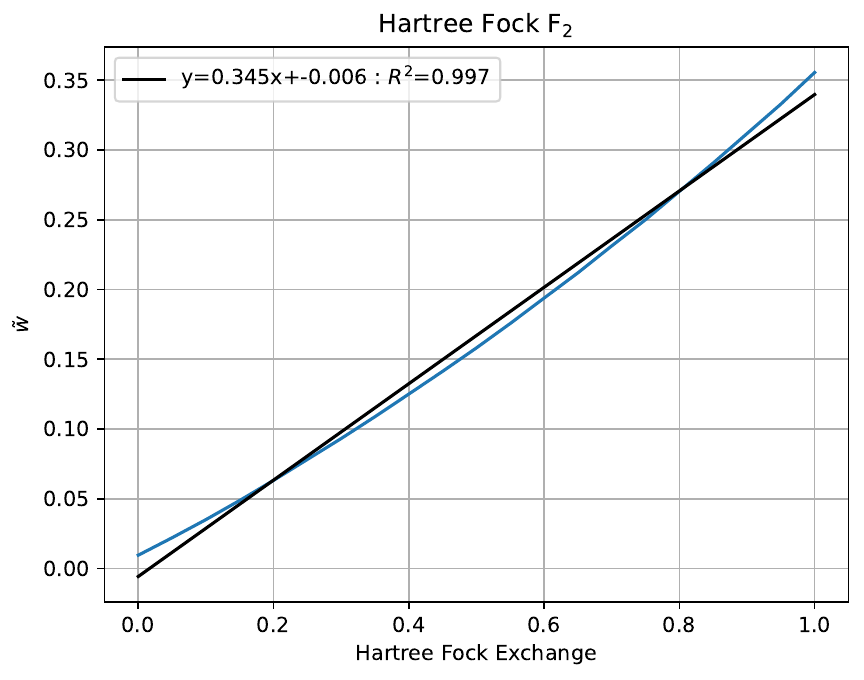}
    \caption{Plot of the fraction of HF exchange in \addtxt{DFA} 1-RDMFT utilizing the HF XC functional vs the optimal $\tilde{w}$ value for F$_2$. The linear regression is also plotted to emphasize the slight but notable non-linear behavior.}
    \label{fig:HvsHF}
\end{figure}

To further elucidate the dependence of $\tilde{w}$ on HF exchange, we linearly interpolate between Hartree and Hartree-Fock theory by modulating the amount of exchange in \addtxt{DFA} 1-RDMFT using the HF functional. As shown in Figure \ref{fig:HvsHF} for F$_2$, we now observe a slight curvature in the relationship between $\tilde{w}$ and HF exchange. This implies that the previously observed linear behavior in XC functionals is due to the interplay between the functional's exchange term and HF exchange with both being slightly non-linear on their own. In the 1-RDM term $\tilde{w}\text{Tr}({}^1D^2-{}^1D)$ in \addtxt{DFA} 1-RDMFT, $\tilde{w}$ has a linear dependence on the exchange integrals; however, ${}^1D^2-{}^1D$ is non-linear. This is likely the origin for the non-linear relationship between $\tilde{w}$ and HF exchange. \\

\section{Conclusion}

In this paper, we explored how the accuracy of the previously developed \addtxt{DFA} 1-RDMFT, a method which combines a 1-RDM functional with existing XC functionals, depends on the utilized XC functional and we compare the results to the traditional UKS-DFT approach. In this \addtxt{DFA} 1-RDMFT framework, the role of the 1-RDM term is to capture strong electron correlation effects, while the XC functional captures the remaining (dynamic) correlation effects. The balance between the two terms is tuned via the functional dependent parameter $\kappa$. The \addtxt{DFA} 1-RDMFT framework needs to smoothly transition between XC-only regimes in the absence of strong correlation, where the KS-DFT result is recovered, and those where the 1-RDM term is non-zero. An XC functional that already yields good results in strongly correlated systems may produce poor results when used in \addtxt{DFA} 1-RDMFT due to overlapping contributions to the energy. \\

Systematically benchmarking 190 functionals available in the LibXC library, we find that functionals on the same rung of Jacob's ladder generally yield comparable performance within either UKS-DFT or \addtxt{DFA} 1-RDMFT. However, the trend of functionals generally increasing in accuracy as one ascends Jacob's ladder, which is observed in UKS-DFT, is reversed in \addtxt{DFA} 1-RDMFT, where GGAs and MGGAs outperformed Hybrid and RSH functionals.  Importantly, \addtxt{DFA} 1-RDMFT yields lower average cumulative errors for every rung of Jacob's ladder except for RSHs, while displaying significantly less dependence on the underlying XC functional as compared to UKS-DFT. In general, we note that \addtxt{DFA} 1-RDMFT displays better convergence behavior, lower overall errors, and less functional dependence than UKS-DFT.\\

Investigating the performance of Hybrid XC functionals in \addtxt{DFA} 1-RDMFT, we also reveal a linear dependence of \addtxt{DFA} 1-RDMFT's $\tilde{w}$ value on the fraction of HF exchange in XC functionals, with a higher HF exchange fraction necessitating a larger $\tilde{w}$ value. While $\tilde{w}$ has a dependence on the exchange integrals, by utilizing this exchange relationship through Hartree and Hartree-Fock theory, we find that the linearity arises from a cancellation of non-linear dependencies in the XC functional's exchange term and HF exchange on $\tilde{w}$.\\ 

Furthermore, we also obtained the $\kappa$ values necessary to utilize the vast majority of functionals available in LibXC within \addtxt{DFA} 1-RDMFT, significantly expanding its utility. These values are tabulated in the SI (Table S2). We find that a low $\kappa$ value does not necessarily yield high accuracy. Instead, we find that functionals with $\kappa$ $\approx$ 0.075 yield the lowest errors, implying that a careful balance between electron correlation recovered through the XC functional and the 1-RDM functional is required to ensure a complete description of electron correlation effects in the presence of strong correlation. The lack of a clear relationship between the \addtxt{DFA} 1-RDMFT $\kappa$ value for a given XC functional and its accuracy in UKS-DFT, implies that \addtxt{DFA} 1-RDMFT and UKS-DFT capture strong correlation effects through fundamentally different mechanisms. \\

\addtxt{Further efforts are currently underway to verify the transferability of the insights gained in this work, including utilizing a subset of the functionals demonstrated to provide high accuracy, to investigate the properties and reactivity of larger, more complex strongly correlated systems of chemical interest. We believe that the overall trends and insights found throughout this work are transferable beyond diatomics. For the systems considered in this work, our data suggests that TPSSh and N12 present widely available XC functionals with good accuracy in both DFT and DFA 1-RDMFT across both single-reference and strongly correlated regimes.}\\

\begin{acknowledgments}
Daniel Gibney thanks Samuel Warren for insightful discussions regarding the HF exchange term.
The authors thank the University of Minnesota for startup funding. 

\end{acknowledgments}

\pagebreak

\section*{Supporting Information}
The Supplemental Material contains the max error figures, the optimal $\tilde{w}$ values for 100\% HF exchange SCAN, B3LYP, TPSS, and PBE, plots of cumulative error vs $\kappa$, and the tabulated $\kappa$, average max and cumulative error values and \addtxt{timing data for N$_2$}.

\bibliographystyle{achemso}

\bibliography{citations.bib}

\end{document}



\title{Benchmarking and contrasting exchange-correlation functional differences in response to static correlation in unrestricted Kohn-Sham and \addtxt{a hybrid} 1-electron reduced density matrix functional theory}

\author{Daniel Gibney}
\author{Jan-Niklas Boyn}%
\affiliation{Department of Chemistry, University of Minnesota, Minneapolis, Minnesota 55455, United States}%
\date{Submitted \today}

\maketitle

\begin{figure}
    \centering
    \includegraphics[width=0.5\linewidth]{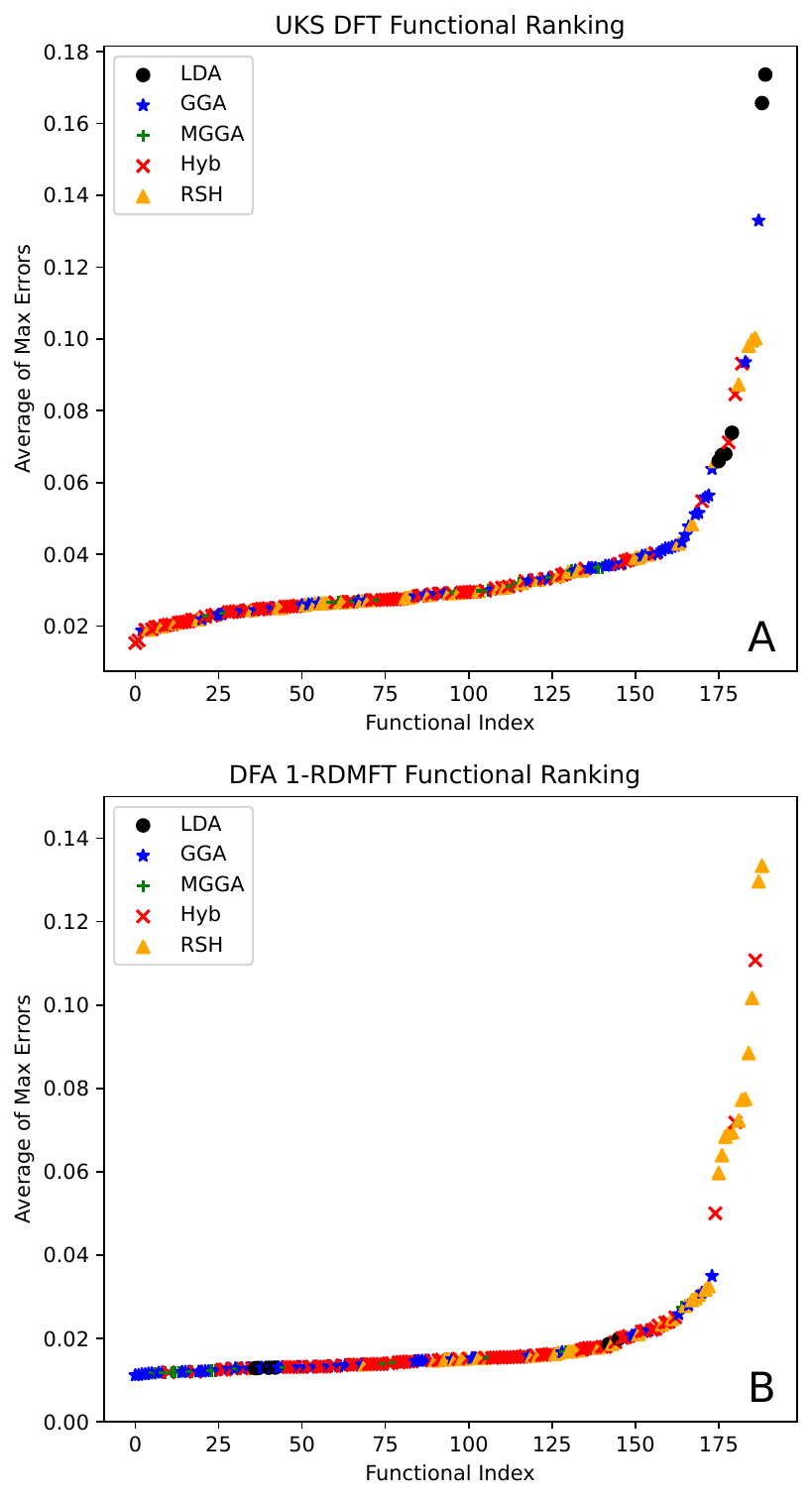}
    \caption{Plots the functional rankings for UKS DFT (A) and \addtxt{DFA} 1-
RDMFT (B). The functionals are ranked in order of increasing
average of max errors.}
    \label{supp-fig:functionals_ranked}
\end{figure}

\begin{figure}
    \centering
    \includegraphics[width=0.5\linewidth]{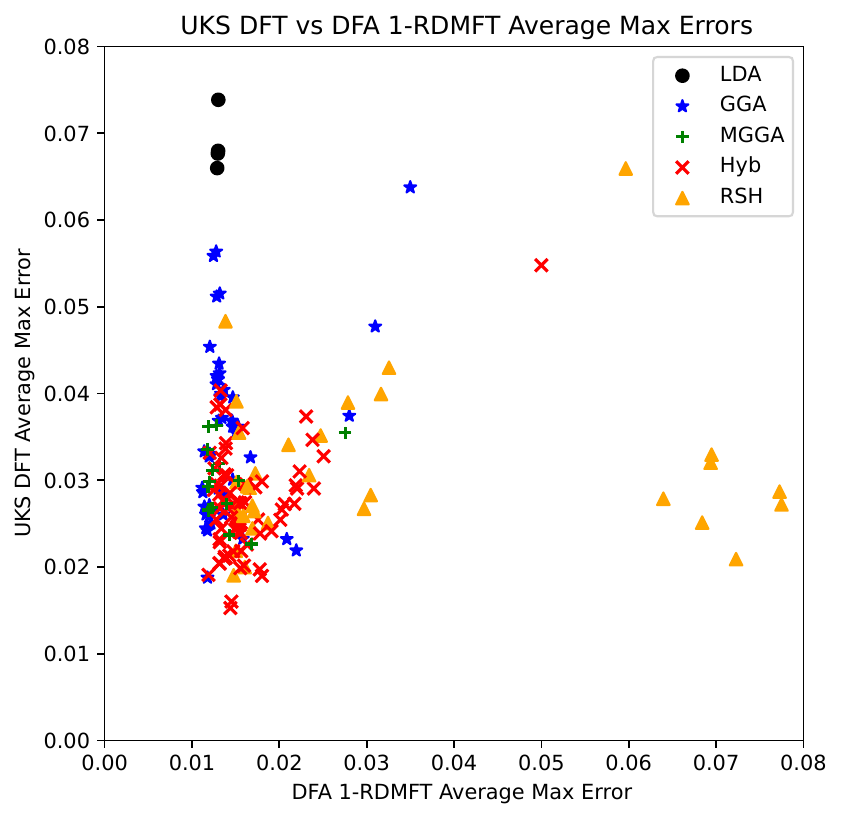}
    \caption{Comparison between average max error from UKS DFT
and \addtxt{DFA} 1-RDMFT for the majority of functionals investigated in
this work.}
    \label{supp-fig:functionals_ranked}
\end{figure}
\begin{figure}
    \centering
    \includegraphics[width=0.5\linewidth]{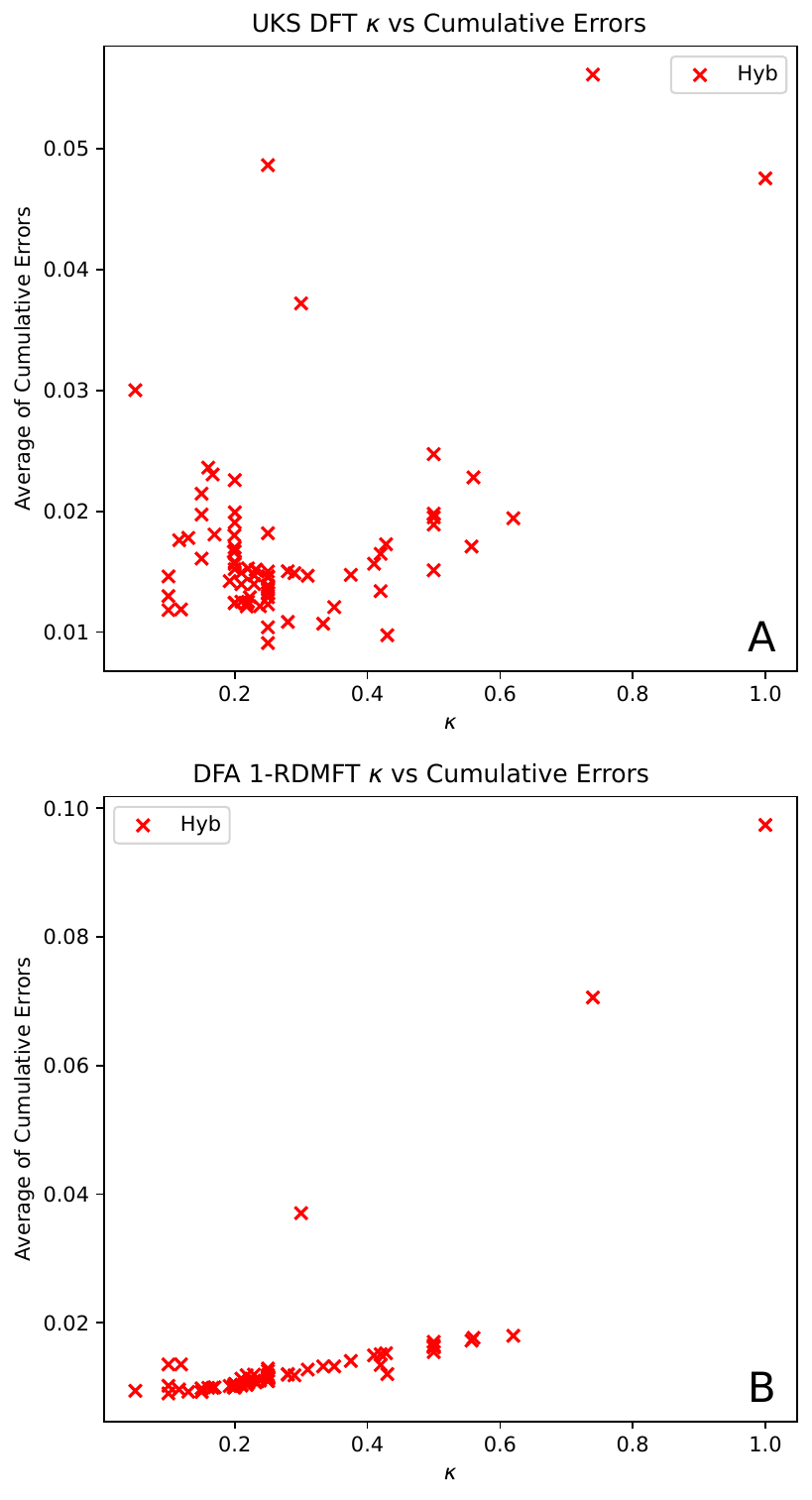}
    \caption{Comparison between average cumulative error from UKS DFT
and \addtxt{DFA} 1-RDMFT with $\kappa$ for the hybrid functionals investigated in
this work.}
    \label{supp-fig:HFvsError}
\end{figure}
\begin{table}[]
    \centering
    \begin{tabular}{l|c|c|c|c}
    System & \multicolumn{4}{c}{Functional}\\
                 & B3LYP                   & PBE                &  SCAN                 & TPSS\\\hline
        C$_2$	 &0.257	                   &0.264	            &0.256	                &0.263\\
        CN	     &0.306	   &0.310	            &0.304	                &0.310\\
        CO	     &0.300	                   &0.305	            &0.298	                &0.304\\
        F$_2$	 &0.366   &0.369	            &0.364	                &0.367\\
        N$_2$	 &0.340	                   &0.345	            &0.338	                &0.345\\
        NO	     &0.360	       &0.366	&0.360	    &0.365\\
        S$_2$	 &0.195	   &0.201	&0.197	&0.201\\
        SiO	     &0.242	       &0.246	&0.240	    &0.245
    \end{tabular}
    \captionsetup{labelformat=empty}
    \caption{TABLE S1: Optimal $\tilde{w}$ values for each functional with 100\% HF exchange.}
    \label{SI:tab:HF_exchange}
\end{table}

\begin{table}[]
    \centering    
    \begin{tabular}{l|c|c|c|c|c|c|c|}
    \multicolumn{2}{c|}{} & \multicolumn{3}{|c|}{\addtxt{DFA 1-RDMFT}} & \multicolumn{3}{|c|}{\addtxt{UKS-DFT}}\\\hline
    \multicolumn{2}{c}{} & \multicolumn{1}{|c|}{} & \multicolumn{2}{|c|}{\addtxt{Average Errors}} & & \multicolumn{2}{|c|}{\addtxt{Average Errors}}\\
Functional     & Optimal $\kappa$ & \addtxt{Functional Index} & \addtxt{Max} & \addtxt{Cumulative} & \addtxt{Functional Index} & \addtxt{Max} & \addtxt{Cumulative}\\ \hline
\verb|gga_xc_mohlyp2|	&0.0741	& \addtxt{134}	& \addtxt{0.0158}	& \addtxt{0.0129}	& \addtxt{28}	& \addtxt{0.0232} 	& \addtxt{0.0129}  \\
\verb|gga_xc_hcth_p76|	&0.0776	& \addtxt{128}	& \addtxt{0.0153}	& \addtxt{0.0124}	& \addtxt{141}	& \addtxt{0.0362} 	& \addtxt{0.0249}  \\
\verb|gga_xc_th_fcfo|	&0.0839	& \addtxt{146}	& \addtxt{0.0219}	& \addtxt{0.0143}	& \addtxt{53}	& \addtxt{0.0219} 	& \addtxt{0.0143}  \\
\verb|gga_xc_mohlyp|	&0.0840	& \addtxt{122}	& \addtxt{0.0153}	& \addtxt{0.012}	& \addtxt{145}	& \addtxt{0.0362} 	& \addtxt{0.0272}  \\
\verb|gga_xc_th2|	&0.0847	& \addtxt{113}	& \addtxt{0.0147}	& \addtxt{0.0116}	& \addtxt{137}	& \addtxt{0.0301} 	& \addtxt{0.0234}  \\
\verb|gga_xc_th3|	&0.0867	& \addtxt{76}	& \addtxt{0.0138}	& \addtxt{0.0106}	& \addtxt{120}	& \addtxt{0.0283} 	& \addtxt{0.0204}  \\
\verb|gga_xc_beefvdw|	&0.0868	& \addtxt{15}	& \addtxt{0.0122}	& \addtxt{0.009}	& \addtxt{107}	& \addtxt{0.025} 	& \addtxt{0.0188}  \\
\verb|gga_xc_xlyp|	&0.0870	& \addtxt{95}	& \addtxt{0.015}	& \addtxt{0.0112}	& \addtxt{153}	& \addtxt{0.0355} 	& \addtxt{0.0296}  \\
\verb|gga_xc_pbelyp1w|	&0.0890	& \addtxt{80}	& \addtxt{0.0147}	& \addtxt{0.0107}	& \addtxt{162}	& \addtxt{0.0395} 	& \addtxt{0.0334}  \\
\verb|gga_xc_ncap|	&0.0890	& \addtxt{106}	& \addtxt{0.0147}	& \addtxt{0.0115}	& \addtxt{152}	& \addtxt{0.0362} 	& \addtxt{0.0296}  \\
\verb|gga_xc_th_fco|	&0.0893	& \addtxt{137}	& \addtxt{0.0209}	& \addtxt{0.0132}	& \addtxt{74}	& \addtxt{0.0232} 	& \addtxt{0.0156}  \\
\verb|gga_xc_th1|	&0.0893	& \addtxt{65}	& \addtxt{0.013}	& \addtxt{0.0104}	& \addtxt{122}	& \addtxt{0.029} 	& \addtxt{0.0206}  \\
\verb|gga_xc_opwlyp_d|	&0.0902	& \addtxt{87}	& \addtxt{0.0146}	& \addtxt{0.011}	& \addtxt{155}	& \addtxt{0.0368} 	& \addtxt{0.0308}  \\
\verb|gga_xc_oblyp_d|	&0.0903	& \addtxt{84}	& \addtxt{0.0145}	& \addtxt{0.0109}	& \addtxt{156}	& \addtxt{0.0369} 	& \addtxt{0.0308}  \\
\verb|gga_x_chachiyogga_c_chachiyo|	&0.0905	& \addtxt{73}	& \addtxt{0.0134}	& \addtxt{0.0105}	& \addtxt{149}	& \addtxt{0.0372} 	& \addtxt{0.0287}  \\
\verb|gga_xc_b97_d|	&0.0915	& \addtxt{27}	& \addtxt{0.0121}	& \addtxt{0.0093}	& \addtxt{114}	& \addtxt{0.0263} 	& \addtxt{0.0196}  \\
\verb|gga_xc_opbe_d|	&0.0918	& \addtxt{63}	& \addtxt{0.0136}	& \addtxt{0.0103}	& \addtxt{158}	& \addtxt{0.0404} 	& \addtxt{0.0321}  \\
\verb|mgga_x_pkzbmgga_c_pkzb|	&0.0920	& \addtxt{68}	& \addtxt{0.0128}	& \addtxt{0.0104}	& \addtxt{146}	& \addtxt{0.0364} 	& \addtxt{0.0278}  \\
\verb|gga_xc_edf1|	&0.0924	& \addtxt{21}	& \addtxt{0.012}	& \addtxt{0.0092}	& \addtxt{142}	& \addtxt{0.033} 	& \addtxt{0.025}  \\
\verb|gga_xc_th4|	&0.0928	& \addtxt{8}	& \addtxt{0.0116}	& \addtxt{0.0087}	& \addtxt{98}	& \addtxt{0.026} 	& \addtxt{0.018}  \\
\verb|mgga_xc_tpsslyp1w|	&0.0931	& \addtxt{72}	& \addtxt{0.014}	& \addtxt{0.0105}	& \addtxt{134}	& \addtxt{0.0273} 	& \addtxt{0.0229}  \\
\verb|gga_xc_hcth_93|	&0.0937	& \addtxt{5}	& \addtxt{0.0114}	& \addtxt{0.0085}	& \addtxt{90}	& \addtxt{0.027} 	& \addtxt{0.0171}  \\
\verb|gga_xc_pbe1w|	&0.0938	& \addtxt{40}	& \addtxt{0.0132}	& \addtxt{0.0098}	& \addtxt{159}	& \addtxt{0.0397} 	& \addtxt{0.0323}  \\
\verb|gga_x_xpbegga_c_xpbe|	&0.0938	& \addtxt{37}	& \addtxt{0.0131}	& \addtxt{0.0097}	& \addtxt{151}	& \addtxt{0.0369} 	& \addtxt{0.0294}  \\
\verb|gga_xc_th_fc|	&0.0939	& \addtxt{168}	& \addtxt{0.035}	& \addtxt{0.0254}	& \addtxt{175}	& \addtxt{0.0638} 	& \addtxt{0.0467}  \\
\verb|gga_x_pbe_molgga_c_pbe_mol|	&0.0940	& \addtxt{39}	& \addtxt{0.0132}	& \addtxt{0.0097}	& \addtxt{164}	& \addtxt{0.0423} 	& \addtxt{0.0336}  \\
\verb|gga_xc_b97_3c|	&0.0942	& \addtxt{23}	& \addtxt{0.012}	& \addtxt{0.0092}	& \addtxt{116}	& \addtxt{0.0271} 	& \addtxt{0.0198}  \\
\verb|gga_x_apbegga_c_apbe|	&0.0948	& \addtxt{38}	& \addtxt{0.0131}	& \addtxt{0.0097}	& \addtxt{168}	& \addtxt{0.0434} 	& \addtxt{0.0344}  \\
\verb|gga_xc_hcth_147|	&0.0958	& \addtxt{2}	& \addtxt{0.0112}	& \addtxt{0.0085}	& \addtxt{126}	& \addtxt{0.0291} 	& \addtxt{0.0208}  \\
\verb|gga_xc_b97_gga1|	&0.0958	& \addtxt{14}	& \addtxt{0.012}	& \addtxt{0.009}	& \addtxt{86}	& \addtxt{0.0258} 	& \addtxt{0.0168}  \\
\verb|gga_xc_hcth_120|	&0.0960	& \addtxt{4}	& \addtxt{0.0112}	& \addtxt{0.0085}	& \addtxt{124}	& \addtxt{0.0286} 	& \addtxt{0.0207}  \\
\verb|gga_xc_hcth_407p|	&0.0961	& \addtxt{6}	& \addtxt{0.012}	& \addtxt{0.0086}	& \addtxt{130}	& \addtxt{0.0327} 	& \addtxt{0.0223}  \\
\verb|gga_xc_mpwlyp1w|	&0.0968	& \addtxt{166}	& \addtxt{0.028}	& \addtxt{0.022}	& \addtxt{157}	& \addtxt{0.0374} 	& \addtxt{0.0312}  \\
\verb|gga_xc_hcth_407|	&0.0969	& \addtxt{3}	& \addtxt{0.0118}	& \addtxt{0.0085}	& \addtxt{63}	& \addtxt{0.0242} 	& \addtxt{0.0148}  \\
\verb|gga_x_pbegga_c_pbe|	&0.0970	& \addtxt{34}	& \addtxt{0.013}	& \addtxt{0.0097}	& \addtxt{166}	& \addtxt{0.0416} 	& \addtxt{0.0337}  \\
\verb|gga_x_pbe_gaussiangga_c_pbe_gaussian|	&0.0970	& \addtxt{35}	& \addtxt{0.013}	& \addtxt{0.0097}	& \addtxt{165}	& \addtxt{0.0416} 	& \addtxt{0.0337}  \\
\verb|gga_x_pw91gga_c_pw91|	&0.0976	& \addtxt{31}	& \addtxt{0.0128}	& \addtxt{0.0096}	& \addtxt{160}	& \addtxt{0.0411} 	& \addtxt{0.0326}  \\
\verb|gga_x_rge2gga_c_rge2|	&0.0991	& \addtxt{42}	& \addtxt{0.0128}	& \addtxt{0.0098}	& \addtxt{161}	& \addtxt{0.042} 	& \addtxt{0.0326}  \\
\verb|mgga_xc_otpss_d|	&0.0999	& \addtxt{147}	& \addtxt{0.0168}	& \addtxt{0.0149}	& \addtxt{80}	& \addtxt{0.0226} 	& \addtxt{0.0159}  \\
\verb|mgga_x_tpssmgga_c_tpss|	&0.1002	& \addtxt{30}	& \addtxt{0.0124}	& \addtxt{0.0095}	& \addtxt{138}	& \addtxt{0.0312} 	& \addtxt{0.0236}  \\
\verb|gga_x_pbeintgga_c_pbeint|	&0.1007	& \addtxt{36}	& \addtxt{0.0128}	& \addtxt{0.0097}	& \addtxt{170}	& \addtxt{0.0512} 	& \addtxt{0.0406}  \\
\verb|gga_x_am05gga_c_am05|	&0.1017	& \addtxt{22}	& \addtxt{0.0125}	& \addtxt{0.0092}	& \addtxt{173}	& \addtxt{0.0559} 	& \addtxt{0.043}  \\
\verb|gga_x_pbefegga_c_pbefe|	&0.1020	& \addtxt{54}	& \addtxt{0.0132}	& \addtxt{0.01}	& \addtxt{172}	& \addtxt{0.0515} 	& \addtxt{0.0424}  \\
\verb|gga_x_gamgga_c_gam|	&0.1020	& \addtxt{11}	& \addtxt{0.0131}	& \addtxt{0.0088}	& \addtxt{118}	& \addtxt{0.0319} 	& \addtxt{0.0198}  \\
\verb|gga_xc_kt1|	&0.1023	& \addtxt{145}	& \addtxt{0.0167}	& \addtxt{0.0143}	& \addtxt{147}	& \addtxt{0.0326} 	& \addtxt{0.0281}  \\
\verb|mgga_x_revtpssmgga_c_revtpss|	&0.1023	& \addtxt{29}	& \addtxt{0.0123}	& \addtxt{0.0094}	& \addtxt{125}	& \addtxt{0.0268} 	& \addtxt{0.0207}  \\
\verb|mgga_x_revtmmgga_c_revtm|	&0.1025	& \addtxt{13}	& \addtxt{0.0119}	& \addtxt{0.0088}	& \addtxt{148}	& \addtxt{0.0362} 	& \addtxt{0.0286}  \\
\verb|mgga_x_tmmgga_c_tm|	&0.1029	& \addtxt{7}	& \addtxt{0.0118}	& \addtxt{0.0087}	& \addtxt{144}	& \addtxt{0.0335} 	& \addtxt{0.0259}  \\
\verb|gga_xc_kt3|	&0.1033	& \addtxt{9}	& \addtxt{0.0116}	& \addtxt{0.0088}	& \addtxt{93}	& \addtxt{0.0245} 	& \addtxt{0.0176}  \\
\verb|gga_x_sg4gga_c_sg4|	&0.1034	& \addtxt{12}	& \addtxt{0.012}	& \addtxt{0.0088}	& \addtxt{167}	& \addtxt{0.0454} 	& \addtxt{0.0338}  \\
\verb|gga_x_n12gga_c_n12|	&0.1035	& \addtxt{19}	& \addtxt{0.0118}	& \addtxt{0.0091}	& \addtxt{12}	& \addtxt{0.0188} 	& \addtxt{0.012}  \\
\verb|gga_x_pbe_solgga_c_pbe_sol|	&0.1040	& \addtxt{32}	& \addtxt{0.0128}	& \addtxt{0.0096}	& \addtxt{174}	& \addtxt{0.0564} 	& \addtxt{0.0445}  \\
\verb|gga_xc_hcth_p14|	&0.1043	& \addtxt{1}	& \addtxt{0.0114}	& \addtxt{0.0084}	& \addtxt{131}	& \addtxt{0.0333} 	& \addtxt{0.0225}  \\
\verb|hyb_gga_xc_mpwlyp1m|	&0.1048	& \addtxt{28}	& \addtxt{0.0128}	& \addtxt{0.0094}	& \addtxt{154}	& \addtxt{0.0384} 	& \addtxt{0.03}  \\
\verb|gga_xc_kt2|	&0.1101	& \addtxt{83}	& \addtxt{0.0136}	& \addtxt{0.0109}	& \addtxt{123}	& \addtxt{0.026} 	& \addtxt{0.0206}  \\
\verb|lda_xc_corrksdt|	&0.1111	& \addtxt{55}	& \addtxt{0.013}	& \addtxt{0.01}	& \addtxt{183}	& \addtxt{0.0739} 	& \addtxt{0.0588}  \\
\verb|lda_xc_gdsmfb|	&0.1111	& \addtxt{53}	& \addtxt{0.013}	& \addtxt{0.01}	& \addtxt{180}	& \addtxt{0.068} 	& \addtxt{0.0535}  \\
\verb|lda_xc_ksdt|	&0.1113	& \addtxt{51}	& \addtxt{0.013}	& \addtxt{0.01}	& \addtxt{179}	& \addtxt{0.0677} 	& \addtxt{0.0526}  \\
\verb|lda_xc_teter93|	&0.1118	& \addtxt{45}	& \addtxt{0.0129}	& \addtxt{0.01}	& \addtxt{178}	& \addtxt{0.066} 	& \addtxt{0.051}  \\
\verb|mgga_x_m06_lmgga_c_m06_l|	&0.1123	& \addtxt{10}	& \addtxt{0.0118}	& \addtxt{0.0088}	& \addtxt{96}	& \addtxt{0.0293} 	& \addtxt{0.0177}  \\
\verb|mgga_x_m11_lmgga_c_m11_l|	&0.1125	& \addtxt{133}	& \addtxt{0.0163}	& \addtxt{0.0128}	& \addtxt{78}	& \addtxt{0.0293} 	& \addtxt{0.0158}  \\
\verb|gga_xc_th_fl|	&0.1147	& \addtxt{74}	& \addtxt{0.0133}	& \addtxt{0.0105}	& \addtxt{150}	& \addtxt{0.04} 	& \addtxt{0.0289}  \\
\verb|mgga_x_r2scan01mgga_c_r2scan01|	&0.1160	& \addtxt{17}	& \addtxt{0.0119}	& \addtxt{0.009}	& \addtxt{88}	& \addtxt{0.0266} 	& \addtxt{0.0169}  \\
\verb|mgga_x_r2scanmgga_c_r2scan|	&0.1162	& \addtxt{18}	& \addtxt{0.0119}	& \addtxt{0.009}	& \addtxt{87}	& \addtxt{0.0265} 	& \addtxt{0.0168}  \\
\verb|mgga_x_rscanmgga_c_rscan|	&0.1164	& \addtxt{20}	& \addtxt{0.012}	& \addtxt{0.0091}	& \addtxt{109}	& \addtxt{0.0298} 	& \addtxt{0.0189}  \\
    \end{tabular}
    \label{tab:my_label}
\end{table}

\begin{table}[]
    \centering    
    \begin{tabular}{l|c|c|c|c|c|c|c|}
    \multicolumn{2}{c|}{} & \multicolumn{3}{|c|}{\addtxt{DFA 1-RDMFT}} & \multicolumn{3}{|c|}{\addtxt{UKS-DFT}}\\\hline
    \multicolumn{2}{c}{} & \multicolumn{1}{|c|}{} & \multicolumn{2}{|c|}{\addtxt{Average Errors}} & & \multicolumn{2}{|c|}{\addtxt{Average Errors}}\\
Functional     & Optimal $\kappa$ & \addtxt{Functional Index} & \addtxt{Max} & \addtxt{Cumulative} & \addtxt{Functional Index} & \addtxt{Max} & \addtxt{Cumulative}\\ \hline
\verb|hyb_mgga_xc_x1b95|	&0.1174	& \addtxt{175}	& \addtxt{0.05}	& \addtxt{0.0371}	& \addtxt{169}	& \addtxt{0.0548} 	& \addtxt{0.0372}  \\
\verb|mgga_xc_hle17|	&0.1259	& \addtxt{161}	& \addtxt{0.0275}	& \addtxt{0.0173}	& \addtxt{128}	& \addtxt{0.0355} 	& \addtxt{0.0215}  \\
\verb|mgga_x_mn15_lmgga_c_mn15_l|	&0.1268	& \addtxt{75}	& \addtxt{0.0143}	& \addtxt{0.0105}	& \addtxt{41}	& \addtxt{0.0237} 	& \addtxt{0.0138}  \\
\verb|lda_xc_lp_a|	&0.1275	& \addtxt{127}	& \addtxt{0.0187}	& \addtxt{0.0124}	& \addtxt{189}	& \addtxt{0.1657} 	& \addtxt{0.1407}  \\
\verb|hyb_gga_xc_o3lyp|	&0.1285	& \addtxt{33}	& \addtxt{0.0129}	& \addtxt{0.0096}	& \addtxt{94}	& \addtxt{0.0295} 	& \addtxt{0.0176}  \\
\verb|gga_x_q2dgga_c_q2d|	&0.1297	& \addtxt{157}	& \addtxt{0.0207}	& \addtxt{0.0167}	& \addtxt{187}	& \addtxt{0.0935} 	& \addtxt{0.0767}  \\
\verb|hyb_mgga_xc_tpssh|	&0.1302	& \addtxt{16}	& \addtxt{0.0119}	& \addtxt{0.009}	& \addtxt{10}	& \addtxt{0.0191} 	& \addtxt{0.0118}  \\
\verb|hyb_gga_xc_whpbe0|	&0.1321	& \addtxt{176}	& \addtxt{0.0597}	& \addtxt{0.0445}	& \addtxt{171}	& \addtxt{0.0659} 	& \addtxt{0.0419}  \\
\verb|mgga_x_revm06_lmgga_c_revm06_l|	&0.1329	& \addtxt{97}	& \addtxt{0.0154}	& \addtxt{0.0113}	& \addtxt{69}	& \addtxt{0.0299} 	& \addtxt{0.0151}  \\
\verb|lda_xc_lp_b|	&0.1336	& \addtxt{139}	& \addtxt{0.0199}	& \addtxt{0.0133}	& \addtxt{190}	& \addtxt{0.1736} 	& \addtxt{0.1427}  \\
\verb|hyb_mgga_xc_revtpssh|	&0.1347	& \addtxt{141}	& \addtxt{0.0156}	& \addtxt{0.0135}	& \addtxt{56}	& \addtxt{0.0239} 	& \addtxt{0.0146}  \\
\verb|hyb_mgga_xc_tpss1kcis|	&0.1364	& \addtxt{25}	& \addtxt{0.0125}	& \addtxt{0.0093}	& \addtxt{97}	& \addtxt{0.0289} 	& \addtxt{0.0178}  \\
\verb|hyb_gga_xc_b3lyps|	&0.1368	& \addtxt{24}	& \addtxt{0.012}	& \addtxt{0.0093}	& \addtxt{127}	& \addtxt{0.0332} 	& \addtxt{0.0215}  \\
\verb|hyb_gga_xc_b97_1p|	&0.1381	& \addtxt{41}	& \addtxt{0.0128}	& \addtxt{0.0098}	& \addtxt{81}	& \addtxt{0.0255} 	& \addtxt{0.0161}  \\
\verb|hyb_mgga_xc_mpw1kcis|	&0.1382	& \addtxt{26}	& \addtxt{0.0126}	& \addtxt{0.0093}	& \addtxt{115}	& \addtxt{0.0314} 	& \addtxt{0.0197}  \\
\verb|gga_xc_hle16|	&0.1400	& \addtxt{169}	& \addtxt{0.031}	& \addtxt{0.0254}	& \addtxt{143}	& \addtxt{0.0477} 	& \addtxt{0.0259}  \\
\verb|hyb_gga_xc_wp04|	&0.1404	& \addtxt{142}	& \addtxt{0.018}	& \addtxt{0.0135}	& \addtxt{11}	& \addtxt{0.0299} 	& \addtxt{0.0119}  \\
\verb|hyb_mgga_xc_r2scanh|	&0.1439	& \addtxt{59}	& \addtxt{0.0139}	& \addtxt{0.0102}	& \addtxt{29}	& \addtxt{0.0211} 	& \addtxt{0.013}  \\
\verb|hyb_gga_xc_edf2|	&0.1474	& \addtxt{47}	& \addtxt{0.0134}	& \addtxt{0.01}	& \addtxt{102}	& \addtxt{0.0326} 	& \addtxt{0.0181}  \\
\verb|hyb_gga_xc_sb98_1c|	&0.1494	& \addtxt{57}	& \addtxt{0.0132}	& \addtxt{0.0102}	& \addtxt{52}	& \addtxt{0.0232} 	& \addtxt{0.0142}  \\
\verb|hyb_gga_xc_b97|	&0.1499	& \addtxt{61}	& \addtxt{0.0132}	& \addtxt{0.0102}	& \addtxt{44}	& \addtxt{0.0229} 	& \addtxt{0.014}  \\
\verb|hyb_gga_xc_b3lyp3|	&0.1504	& \addtxt{49}	& \addtxt{0.0132}	& \addtxt{0.01}	& \addtxt{20}	& \addtxt{0.0204} 	& \addtxt{0.0124}  \\
\verb|hyb_gga_xc_b3lyp5|	&0.1505	& \addtxt{50}	& \addtxt{0.0132}	& \addtxt{0.01}	& \addtxt{19}	& \addtxt{0.0204} 	& \addtxt{0.0124}  \\
\verb|hyb_gga_xc_hpbeint|	&0.1507	& \addtxt{44}	& \addtxt{0.0133}	& \addtxt{0.0099}	& \addtxt{136}	& \addtxt{0.0387} 	& \addtxt{0.023}  \\
\verb|hyb_gga_xc_b3lyp|	&0.1507	& \addtxt{52}	& \addtxt{0.0132}	& \addtxt{0.01}	& \addtxt{75}	& \addtxt{0.0267} 	& \addtxt{0.0156}  \\
\verb|hyb_gga_xc_b1wc|	&0.1508	& \addtxt{43}	& \addtxt{0.0133}	& \addtxt{0.0099}	& \addtxt{139}	& \addtxt{0.0403} 	& \addtxt{0.0236}  \\
\verb|hyb_gga_xc_sb98_1b|	&0.1516	& \addtxt{58}	& \addtxt{0.0137}	& \addtxt{0.0102}	& \addtxt{99}	& \addtxt{0.0307} 	& \addtxt{0.018}  \\
\verb|hyb_gga_xc_revb3lyp|	&0.1523	& \addtxt{56}	& \addtxt{0.0133}	& \addtxt{0.01}	& \addtxt{89}	& \addtxt{0.0294} 	& \addtxt{0.017}  \\
\verb|hyb_gga_xc_hse06|	&0.1527	& \addtxt{92}	& \addtxt{0.015}	& \addtxt{0.0111}	& \addtxt{49}	& \addtxt{0.0264} 	& \addtxt{0.0141}  \\
\verb|hyb_gga_xc_b3lyp_mcm1|	&0.1528	& \addtxt{46}	& \addtxt{0.0131}	& \addtxt{0.01}	& \addtxt{84}	& \addtxt{0.0283} 	& \addtxt{0.0167}  \\
\verb|hyb_gga_xc_hjs_pbe|	&0.1530	& \addtxt{94}	& \addtxt{0.015}	& \addtxt{0.0112}	& \addtxt{9}	& \addtxt{0.0219} 	& \addtxt{0.0114}  \\
\verb|hyb_gga_xc_hse03|	&0.1539	& \addtxt{93}	& \addtxt{0.0152}	& \addtxt{0.0112}	& \addtxt{48}	& \addtxt{0.0265} 	& \addtxt{0.014}  \\
\verb|hyb_gga_xc_mb3lyp_rc04|	&0.1544	& \addtxt{62}	& \addtxt{0.0138}	& \addtxt{0.0103}	& \addtxt{119}	& \addtxt{0.0336} 	& \addtxt{0.0199}  \\
\verb|hyb_gga_xc_b97_1|	&0.1554	& \addtxt{67}	& \addtxt{0.0134}	& \addtxt{0.0104}	& \addtxt{65}	& \addtxt{0.0244} 	& \addtxt{0.0149}  \\
\verb|hyb_gga_xc_b3pw91|	&0.1555	& \addtxt{71}	& \addtxt{0.0139}	& \addtxt{0.0105}	& \addtxt{72}	& \addtxt{0.0276} 	& \addtxt{0.0152}  \\
\verb|hyb_gga_xc_b3p86|	&0.1559	& \addtxt{60}	& \addtxt{0.0136}	& \addtxt{0.0102}	& \addtxt{79}	& \addtxt{0.0276} 	& \addtxt{0.0158}  \\
\verb|hyb_gga_xc_hjs_b97x|	&0.1560	& \addtxt{109}	& \addtxt{0.0151}	& \addtxt{0.0115}	& \addtxt{82}	& \addtxt{0.0298} 	& \addtxt{0.0162}  \\
\verb|hyb_gga_xc_hapbe|	&0.1567	& \addtxt{69}	& \addtxt{0.0139}	& \addtxt{0.0104}	& \addtxt{132}	& \addtxt{0.0381} 	& \addtxt{0.0226}  \\
\verb|hyb_gga_xc_x3lyp|	&0.1569	& \addtxt{64}	& \addtxt{0.0137}	& \addtxt{0.0103}	& \addtxt{14}	& \addtxt{0.021} 	& \addtxt{0.0121}  \\
\verb|hyb_gga_xc_mpw3pw|	&0.1571	& \addtxt{70}	& \addtxt{0.014}	& \addtxt{0.0105}	& \addtxt{85}	& \addtxt{0.0306} 	& \addtxt{0.0167}  \\
\verb|hyb_gga_xc_b3p86_nwchem|	&0.1576	& \addtxt{66}	& \addtxt{0.0139}	& \addtxt{0.0104}	& \addtxt{110}	& \addtxt{0.0343} 	& \addtxt{0.0191}  \\
\verb|hyb_gga_xc_sb98_2c|	&0.1580	& \addtxt{81}	& \addtxt{0.0138}	& \addtxt{0.0108}	& \addtxt{21}	& \addtxt{0.0212} 	& \addtxt{0.0125}  \\
\verb|hyb_gga_xc_b97_2|	&0.1584	& \addtxt{101}	& \addtxt{0.0149}	& \addtxt{0.0113}	& \addtxt{22}	& \addtxt{0.0243} 	& \addtxt{0.0125}  \\
\verb|hyb_gga_xc_hjs_pbe_sol|	&0.1588	& \addtxt{98}	& \addtxt{0.0154}	& \addtxt{0.0113}	& \addtxt{103}	& \addtxt{0.0355} 	& \addtxt{0.0181}  \\
\verb|hyb_gga_xc_hse_sol|	&0.1588	& \addtxt{99}	& \addtxt{0.0154}	& \addtxt{0.0113}	& \addtxt{100}	& \addtxt{0.0355} 	& \addtxt{0.018}  \\
\verb|hyb_gga_xc_mpw3lyp|	&0.1589	& \addtxt{117}	& \addtxt{0.0156}	& \addtxt{0.0118}	& \addtxt{17}	& \addtxt{0.0218} 	& \addtxt{0.0123}  \\
\verb|hyb_mgga_xc_pbe1kcis|	&0.1602	& \addtxt{77}	& \addtxt{0.0144}	& \addtxt{0.0106}	& \addtxt{73}	& \addtxt{0.0285} 	& \addtxt{0.0153}  \\
\verb|hyb_gga_xc_sb98_2a|	&0.1621	& \addtxt{82}	& \addtxt{0.0141}	& \addtxt{0.0108}	& \addtxt{71}	& \addtxt{0.0268} 	& \addtxt{0.0152}  \\
\verb|hyb_gga_xc_b3lyp_mcm2|	&0.1627	& \addtxt{100}	& \addtxt{0.0154}	& \addtxt{0.0113}	& \addtxt{26}	& \addtxt{0.0248} 	& \addtxt{0.0129}  \\
\verb|hyb_gga_xc_sb98_2b|	&0.1635	& \addtxt{90}	& \addtxt{0.0143}	& \addtxt{0.011}	& \addtxt{15}	& \addtxt{0.0213} 	& \addtxt{0.0121}  \\
\verb|hyb_gga_xc_sb98_1a|	&0.1643	& \addtxt{120}	& \addtxt{0.0161}	& \addtxt{0.0119}	& \addtxt{46}	& \addtxt{0.0275} 	& \addtxt{0.014}  \\
\verb|hyb_gga_xc_b1lyp|	&0.1645	& \addtxt{86}	& \addtxt{0.0145}	& \addtxt{0.0109}	& \addtxt{1}	& \addtxt{0.016} 	& \addtxt{0.0091}  \\
\verb|gga_x_hcth_agga_c_hcth_a|	&0.1649	& \addtxt{165}	& \addtxt{0.0258}	& \addtxt{0.0201}	& \addtxt{188}	& \addtxt{0.133} 	& \addtxt{0.0936}  \\
\verb|hyb_gga_xc_hse12s|	&0.1652	& \addtxt{143}	& \addtxt{0.0187}	& \addtxt{0.0139}	& \addtxt{77}	& \addtxt{0.0251} 	& \addtxt{0.0158}  \\
\verb|hyb_gga_xc_mpw1lyp|	&0.1658	& \addtxt{88}	& \addtxt{0.0147}	& \addtxt{0.011}	& \addtxt{16}	& \addtxt{0.0218} 	& \addtxt{0.0123}  \\
\verb|hyb_gga_xc_apf|	&0.1658	& \addtxt{89}	& \addtxt{0.0148}	& \addtxt{0.011}	& \addtxt{60}	& \addtxt{0.0273} 	& \addtxt{0.0147}  \\
\verb|hyb_gga_xc_camy_pbeh|	&0.1676	& \addtxt{149}	& \addtxt{0.0179}	& \addtxt{0.015}	& \addtxt{185}	& \addtxt{0.1002} 	& \addtxt{0.0636}  \\
\verb|hyb_gga_xc_relpbe0|	&0.1688	& \addtxt{79}	& \addtxt{0.0145}	& \addtxt{0.0107}	& \addtxt{70}	& \addtxt{0.0257} 	& \addtxt{0.0151}  \\
\verb|hyb_gga_xc_hse12|	&0.1694	& \addtxt{130}	& \addtxt{0.0169}	& \addtxt{0.0125}	& \addtxt{36}	& \addtxt{0.0244} 	& \addtxt{0.0134}  \\
\verb|hyb_gga_xc_case21|	&0.1700	& \addtxt{96}	& \addtxt{0.0152}	& \addtxt{0.0112}	& \addtxt{55}	& \addtxt{0.0274} 	& \addtxt{0.0146}  \\
\verb|hyb_gga_xc_cap0|	&0.1703	& \addtxt{115}	& \addtxt{0.0151}	& \addtxt{0.0117}	& \addtxt{43}	& \addtxt{0.0246} 	& \addtxt{0.0139}  \\
\verb|hyb_gga_xc_pbe_molb0|	&0.1707	& \addtxt{103}	& \addtxt{0.0154}	& \addtxt{0.0114}	& \addtxt{32}	& \addtxt{0.025} 	& \addtxt{0.0132}  \\
\verb|hyb_gga_xc_b1pw91|	&0.1707	& \addtxt{114}	& \addtxt{0.0154}	& \addtxt{0.0116}	& \addtxt{30}	& \addtxt{0.0241} 	& \addtxt{0.0132}  \\
\verb|hyb_gga_xc_pbe_mol0|	&0.1708	& \addtxt{108}	& \addtxt{0.0154}	& \addtxt{0.0115}	& \addtxt{33}	& \addtxt{0.0248} 	& \addtxt{0.0133}  \\
    \end{tabular}
    \label{tab:my_label}
\end{table}

\begin{table}[]
    \centering    
    \begin{tabular}{l|c|c|c|c|c|c|c|}
    \multicolumn{2}{c|}{} & \multicolumn{3}{|c|}{\addtxt{DFA 1-RDMFT}} & \multicolumn{3}{|c|}{\addtxt{UKS-DFT}}\\\hline
    \multicolumn{2}{c}{} & \multicolumn{1}{|c|}{} & \multicolumn{2}{|c|}{\addtxt{Average Errors}} & & \multicolumn{2}{|c|}{\addtxt{Average Errors}}\\
Functional     & Optimal $\kappa$ & \addtxt{Functional Index} & \addtxt{Max} & \addtxt{Cumulative} & \addtxt{Functional Index} & \addtxt{Max} & \addtxt{Cumulative}\\ \hline
\verb|hyb_gga_xc_apbe0|	&0.1714	& \addtxt{107}	& \addtxt{0.0155}	& \addtxt{0.0115}	& \addtxt{38}	& \addtxt{0.0255} 	& \addtxt{0.0136}  \\
\verb|hyb_gga_xc_mpw1pw|	&0.1721	& \addtxt{111}	& \addtxt{0.0156}	& \addtxt{0.0116}	& \addtxt{37}	& \addtxt{0.0251} 	& \addtxt{0.0135}  \\
\verb|hyb_gga_xc_mpw1pbe|	&0.1724	& \addtxt{112}	& \addtxt{0.0155}	& \addtxt{0.0116}	& \addtxt{5}	& \addtxt{0.0199} 	& \addtxt{0.0104}  \\
\verb|hyb_gga_xc_pbeb0|	&0.1726	& \addtxt{104}	& \addtxt{0.0155}	& \addtxt{0.0114}	& \addtxt{57}	& \addtxt{0.0273} 	& \addtxt{0.0146}  \\
\verb|hyb_gga_xc_pbeh|	&0.1731	& \addtxt{105}	& \addtxt{0.0155}	& \addtxt{0.0114}	& \addtxt{54}	& \addtxt{0.0274} 	& \addtxt{0.0145}  \\
\verb|hyb_mgga_xc_b86b95|	&0.1746	& \addtxt{119}	& \addtxt{0.0161}	& \addtxt{0.0119}	& \addtxt{67}	& \addtxt{0.0296} 	& \addtxt{0.015}  \\
\verb|hyb_mgga_xc_pw6b95|	&0.1751	& \addtxt{124}	& \addtxt{0.0163}	& \addtxt{0.012}	& \addtxt{8}	& \addtxt{0.0226} 	& \addtxt{0.0108}  \\
\verb|hyb_mgga_xc_tpss0|	&0.1753	& \addtxt{118}	& \addtxt{0.016}	& \addtxt{0.0119}	& \addtxt{27}	& \addtxt{0.0202} 	& \addtxt{0.0129}  \\
\verb|hyb_mgga_xc_pw86b95|	&0.1768	& \addtxt{116}	& \addtxt{0.0162}	& \addtxt{0.0118}	& \addtxt{64}	& \addtxt{0.0295} 	& \addtxt{0.0149}  \\
\verb|hyb_gga_xc_pbe_sol0|	&0.1781	& \addtxt{102}	& \addtxt{0.0158}	& \addtxt{0.0114}	& \addtxt{104}	& \addtxt{0.036} 	& \addtxt{0.0182}  \\
\verb|hyb_gga_xc_mcam_b3lyp|	&0.1838	& \addtxt{85}	& \addtxt{0.0148}	& \addtxt{0.0109}	& \addtxt{6}	& \addtxt{0.019} 	& \addtxt{0.0104}  \\
\verb|hyb_mgga_xc_r2scan0|	&0.1856	& \addtxt{135}	& \addtxt{0.0178}	& \addtxt{0.0129}	& \addtxt{66}	& \addtxt{0.0239} 	& \addtxt{0.015}  \\
\verb|hyb_mgga_xc_mpw1b95|	&0.1858	& \addtxt{132}	& \addtxt{0.0173}	& \addtxt{0.0127}	& \addtxt{58}	& \addtxt{0.0292} 	& \addtxt{0.0147}  \\
\verb|hyb_gga_xc_cam_pbeh|	&0.1874	& \addtxt{153}	& \addtxt{0.021}	& \addtxt{0.0157}	& \addtxt{129}	& \addtxt{0.0341} 	& \addtxt{0.0218}  \\
\verb|hyb_mgga_xc_b0kcis|	&0.1889	& \addtxt{131}	& \addtxt{0.0178}	& \addtxt{0.0125}	& \addtxt{177}	& \addtxt{0.0846} 	& \addtxt{0.0486}  \\
\verb|hyb_mgga_xc_xb1k|	&0.1912	& \addtxt{123}	& \addtxt{0.0144}	& \addtxt{0.012}	& \addtxt{2}	& \addtxt{0.0153} 	& \addtxt{0.0097}  \\
\verb|hyb_gga_xc_blyp35|	&0.1952	& \addtxt{136}	& \addtxt{0.0178}	& \addtxt{0.0132}	& \addtxt{13}	& \addtxt{0.0197} 	& \addtxt{0.0121}  \\
\verb|hyb_gga_xc_pbe0_13|	&0.1984	& \addtxt{138}	& \addtxt{0.018}	& \addtxt{0.0132}	& \addtxt{7}	& \addtxt{0.019} 	& \addtxt{0.0107}  \\
\verb|hyb_gga_xc_wc04|	&0.2007	& \addtxt{184}	& \addtxt{0.0718}	& \addtxt{0.0706}	& \addtxt{181}	& \addtxt{0.0931} 	& \addtxt{0.0561}  \\
\verb|hyb_gga_xc_camh_b3lyp|	&0.2081	& \addtxt{110}	& \addtxt{0.016}	& \addtxt{0.0115}	& \addtxt{3}	& \addtxt{0.02} 	& \addtxt{0.0099}  \\
\verb|hyb_gga_xc_pbe38|	&0.2109	& \addtxt{144}	& \addtxt{0.0191}	& \addtxt{0.0141}	& \addtxt{61}	& \addtxt{0.0241} 	& \addtxt{0.0147}  \\
\verb|hyb_gga_xc_b97_k|	&0.2114	& \addtxt{140}	& \addtxt{0.0176}	& \addtxt{0.0134}	& \addtxt{35}	& \addtxt{0.0255} 	& \addtxt{0.0134}  \\
\verb|hyb_mgga_xc_mpwkcis1k|	&0.2174	& \addtxt{148}	& \addtxt{0.0201}	& \addtxt{0.0149}	& \addtxt{76}	& \addtxt{0.0254} 	& \addtxt{0.0157}  \\
\verb|hyb_mgga_xc_bb1k|	&0.2182	& \addtxt{150}	& \addtxt{0.0203}	& \addtxt{0.0151}	& \addtxt{83}	& \addtxt{0.0266} 	& \addtxt{0.0165}  \\
\verb|hyb_gga_xc_tuned_cam_b3lyp|	&0.2200	& \addtxt{48}	& \addtxt{0.0138}	& \addtxt{0.01}	& \addtxt{163}	& \addtxt{0.0483} 	& \addtxt{0.0336}  \\
\verb|hyb_gga_xc_mpw1k|	&0.2262	& \addtxt{151}	& \addtxt{0.0206}	& \addtxt{0.0153}	& \addtxt{92}	& \addtxt{0.0272} 	& \addtxt{0.0173}  \\
\verb|hyb_gga_xc_wb97x_d3|	&0.2347	& \addtxt{181}	& \addtxt{0.0723}	& \addtxt{0.0662}	& \addtxt{4}	& \addtxt{0.0209} 	& \addtxt{0.01}  \\
\verb|hyb_gga_xc_cam_b3lyp|	&0.2383	& \addtxt{125}	& \addtxt{0.0171}	& \addtxt{0.0122}	& \addtxt{23}	& \addtxt{0.0265} 	& \addtxt{0.0126}  \\
\verb|hyb_gga_xc_bhandhlyp|	&0.2409	& \addtxt{154}	& \addtxt{0.0219}	& \addtxt{0.0162}	& \addtxt{113}	& \addtxt{0.0294} 	& \addtxt{0.0195}  \\
\verb|hyb_gga_xc_b5050lyp|	&0.2423	& \addtxt{155}	& \addtxt{0.022}	& \addtxt{0.0163}	& \addtxt{108}	& \addtxt{0.029} 	& \addtxt{0.0189}  \\
\verb|hyb_gga_xc_pbe50|	&0.2487	& \addtxt{156}	& \addtxt{0.0223}	& \addtxt{0.0165}	& \addtxt{117}	& \addtxt{0.031} 	& \addtxt{0.0198}  \\
\verb|hyb_mgga_xc_r2scan50|	&0.2541	& \addtxt{158}	& \addtxt{0.0231}	& \addtxt{0.017}	& \addtxt{140}	& \addtxt{0.0373} 	& \addtxt{0.0247}  \\
\verb|hyb_gga_xc_bhandh|	&0.2554	& \addtxt{152}	& \addtxt{0.0217}	& \addtxt{0.0154}	& \addtxt{68}	& \addtxt{0.0273} 	& \addtxt{0.0151}  \\
\verb|hyb_gga_xc_lc_wpbe08_whs|	&0.2558	& \addtxt{186}	& \addtxt{0.1017}	& \addtxt{0.0919}	& \addtxt{59}	& \addtxt{0.0267} 	& \addtxt{0.0147}  \\
\verb|hyb_gga_xc_lc_wpbeh_whs|	&0.2572	& \addtxt{188}	& \addtxt{0.1296}	& \addtxt{0.1171}	& \addtxt{105}	& \addtxt{0.0279} 	& \addtxt{0.0184}  \\
\verb|hyb_gga_xc_cam_qtp_01|	&0.2592	& \addtxt{183}	& \addtxt{0.0775}	& \addtxt{0.0692}	& \addtxt{25}	& \addtxt{0.0272} 	& \addtxt{0.0129}  \\
\verb|hyb_gga_xc_lc_wpbe_whs|	&0.2599	& \addtxt{182}	& \addtxt{0.0773}	& \addtxt{0.0691}	& \addtxt{51}	& \addtxt{0.0287} 	& \addtxt{0.0142}  \\
\verb|hyb_gga_xc_camy_blyp|	&0.2623	& \addtxt{189}	& \addtxt{0.1334}	& \addtxt{0.1205}	& \addtxt{182}	& \addtxt{0.0873} 	& \addtxt{0.0563}  \\
\verb|hyb_gga_xc_wb97x_d|	&0.2637	& \addtxt{126}	& \addtxt{0.0169}	& \addtxt{0.0123}	& \addtxt{40}	& \addtxt{0.0271} 	& \addtxt{0.0137}  \\
\verb|hyb_gga_xc_lrc_wpbeh|	&0.2644	& \addtxt{121}	& \addtxt{0.0166}	& \addtxt{0.012}	& \addtxt{50}	& \addtxt{0.0291} 	& \addtxt{0.0142}  \\
\verb|hyb_gga_xc_lc_blypr|	&0.2644	& \addtxt{170}	& \addtxt{0.0279}	& \addtxt{0.0259}	& \addtxt{95}	& \addtxt{0.039} 	& \addtxt{0.0176}  \\
\verb|hyb_gga_xc_lc_pbeop|	&0.2650	& \addtxt{173}	& \addtxt{0.0316}	& \addtxt{0.031}	& \addtxt{101}	& \addtxt{0.0399} 	& \addtxt{0.018}  \\
\verb|hyb_gga_xc_pbe_2x|	&0.2668	& \addtxt{162}	& \addtxt{0.0238}	& \addtxt{0.0177}	& \addtxt{133}	& \addtxt{0.0347} 	& \addtxt{0.0228}  \\
\verb|hyb_gga_xc_lc_blyp|	&0.2678	& \addtxt{174}	& \addtxt{0.0326}	& \addtxt{0.0324}	& \addtxt{121}	& \addtxt{0.043} 	& \addtxt{0.0205}  \\
\verb|hyb_gga_xc_kmlyp|	&0.2692	& \addtxt{159}	& \addtxt{0.024}	& \addtxt{0.0172}	& \addtxt{91}	& \addtxt{0.0291} 	& \addtxt{0.0171}  \\
\verb|hyb_gga_xc_wb97x|	&0.2729	& \addtxt{172}	& \addtxt{0.0297}	& \addtxt{0.0279}	& \addtxt{34}	& \addtxt{0.0267} 	& \addtxt{0.0133}  \\
\verb|hyb_mgga_xc_lc_tmlyp|	&0.2730	& \addtxt{91}	& \addtxt{0.0159}	& \addtxt{0.0111}	& \addtxt{18}	& \addtxt{0.0259} 	& \addtxt{0.0123}  \\
\verb|hyb_gga_xc_lrc_wpbe|	&0.2762	& \addtxt{78}	& \addtxt{0.0151}	& \addtxt{0.0107}	& \addtxt{106}	& \addtxt{0.0391} 	& \addtxt{0.0187}  \\
\verb|hyb_gga_xc_lc_wpbe|	&0.2808	& \addtxt{177}	& \addtxt{0.064}	& \addtxt{0.0535}	& \addtxt{42}	& \addtxt{0.0279} 	& \addtxt{0.0138}  \\
\verb|hyb_gga_xc_rcam_b3lyp|	&0.2811	& \addtxt{185}	& \addtxt{0.0885}	& \addtxt{0.08}	& \addtxt{24}	& \addtxt{0.0263} 	& \addtxt{0.0128}  \\
\verb|hyb_gga_xc_lcy_pbe|	&0.2839	& \addtxt{190}	& \addtxt{0.1956}	& \addtxt{0.1653}	& \addtxt{186}	& \addtxt{0.0996} 	& \addtxt{0.0645}  \\
\verb|hyb_gga_xc_wb97|	&0.2857	& \addtxt{171}	& \addtxt{0.0305}	& \addtxt{0.0276}	& \addtxt{39}	& \addtxt{0.0283} 	& \addtxt{0.0137}  \\
\verb|hyb_gga_xc_qtp17|	&0.2877	& \addtxt{163}	& \addtxt{0.0251}	& \addtxt{0.018}	& \addtxt{112}	& \addtxt{0.0328} 	& \addtxt{0.0194}  \\
\verb|hyb_gga_xc_cam_qtp_02|	&0.2899	& \addtxt{178}	& \addtxt{0.0684}	& \addtxt{0.0567}	& \addtxt{31}	& \addtxt{0.0251} 	& \addtxt{0.0132}  \\
\verb|hyb_gga_xc_lc_bop|	&0.2934	& \addtxt{180}	& \addtxt{0.0694}	& \addtxt{0.0579}	& \addtxt{62}	& \addtxt{0.032} 	& \addtxt{0.0147}  \\
\verb|hyb_gga_xc_lc_qtp|	&0.2975	& \addtxt{179}	& \addtxt{0.0695}	& \addtxt{0.0575}	& \addtxt{47}	& \addtxt{0.033} 	& \addtxt{0.014}  \\
\verb|hyb_gga_xc_lb07|	&0.3190	& \addtxt{129}	& \addtxt{0.0173}	& \addtxt{0.0124}	& \addtxt{45}	& \addtxt{0.0308} 	& \addtxt{0.014}  \\
\verb|hyb_gga_xc_cam_qtp_00|	&0.3208	& \addtxt{164}	& \addtxt{0.0247}	& \addtxt{0.0181}	& \addtxt{135}	& \addtxt{0.0352} 	& \addtxt{0.023}  \\
\verb|hyb_gga_xc_hflyp|	&0.3336	& \addtxt{187}	& \addtxt{0.1107}	& \addtxt{0.0974}	& \addtxt{176}	& \addtxt{0.0712} 	& \addtxt{0.0475}  \\
\verb|hyb_gga_xc_lc_wpbesol_whs|	&0.3515	& \addtxt{160}	& \addtxt{0.0234}	& \addtxt{0.0172}	& \addtxt{111}	& \addtxt{0.0306} 	& \addtxt{0.0194}  \\
\verb|hyb_gga_xc_lcy_blyp|	&0.3886	& \addtxt{167}	& \addtxt{0.0292}	& \addtxt{0.023}	& \addtxt{184}	& \addtxt{0.0981} 	& \addtxt{0.06}  \\
    \end{tabular}
    \captionsetup{labelformat=empty}
    \caption{TABLE S2: Optimal $\kappa$ values \addtxt{and related functional indices and errors} obtained in this work.}
    \label{tab:my_label}
\end{table}

\clearpage

\begin{table}[h]
\centering
\begin{tabular}{| l | r | c | r | r | c | r |}
\hline
 & \multicolumn{3}{c|}{N2 1.1 \AA} & \multicolumn{3}{c|}{N2 3.0 \AA} \\
\hline
 & PBE & PBE 1-RDMFT & Ratio & PBE & PBE 1-RDMFT & Ratio \\
\hline
dz & 0.040 & 0.155 & 3.886 & 0.040 & 0.167 & 4.167 \\
\hline
tz & 0.062 & 0.289 & 4.682 & 0.061 & 0.300 & 4.900 \\
\hline
qz & 0.122 & 0.626 & 5.143 & 0.129 & 0.617 & 4.791 \\
\hline
5z & 0.312 & 1.330 & 4.267 & 0.314 & 1.298 & 4.139 \\
\hline
\end{tabular}
\captionsetup{labelformat=empty}
    \caption{\addtxt{TABLE S3: Average time, in seconds, for each SCF iteration from UKS PBE and PBE 1-RDMFT in various basis sets and their ratio, considering N$_2$ in varying sized basis sets at 1.1 \AA~and 3.0 \AA. All PBE 1-RDMFT calculations required 6 iterations to converge the energy to $<$ 1e-8. PBE UKS calculations at 1.1 \AA~required 6 cycles to converge and 8 cycles at 3.0 \AA}}
\end{table}

\begin{figure}[h]
    \centering
    \includegraphics[width=1\linewidth]{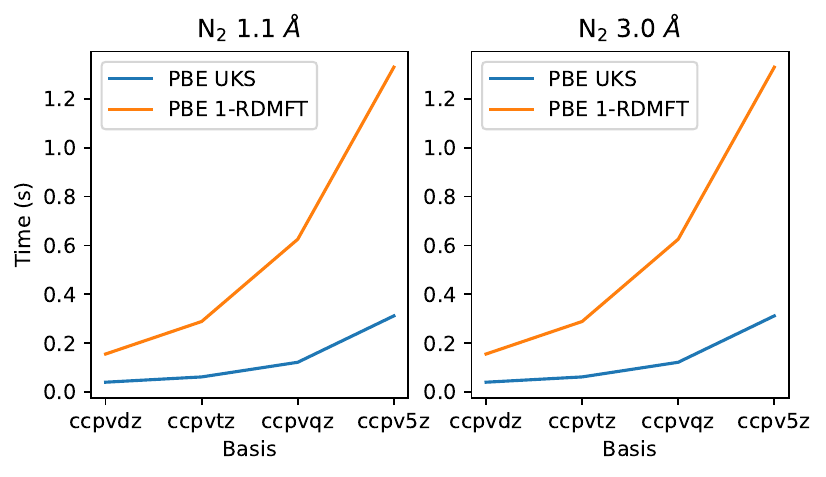}
    \caption{\addtxt{Timing comparisons between PBE UKS and PBE 1-RDMFT with N$_2$ at 1.1 and 3.0 interatomic distances in the cc-pvdz, cc-pvtz, cc-pvqz, and cc-pv5z basis sets.}}
    \label{fig:enter-label}
\end{figure}

\addtxt{Table S3 and Figure S4 show example timing data from UKS DFT and DFA 1-RDMFT in a variety of basis sets using the PySCF package. These results demonstrate comparable scaling with basis set size across UKS-DFT and DFA 1-RDMFT, with an additional prefactor arising from the SDP in the case of the latter. We note that in strongly correlated systems the DFA 1-RDMFT framework tends to converge in fewer cycles than UKS DFT. This is evident in these examples: PBE 1-RDMFT converged after six cycles, irrespective of the basis set used or the bond length considered, while UKS PBE required 6 cycles to converge for all basis sets at the 1.1 \AA~ bond length, but required 8 cycles at the 3.0 \AA~ bond length.}
